\documentclass{vgtc}                          % final (conference style)
\ifpdf%                                % if we use pdflatex
  \pdfoutput=1\relax                   % create PDFs from pdfLaTeX
  \usepackage{graphicx}                % allow us to embed graphics files
  \DeclareGraphicsExtensions{.pdf,.png,.jpg,.jpeg} % for pdflatex we expect .pdf, .png, or .jpg files
\else%                                 % else we use pure latex
  \usepackage{graphicx}                % allow us to embed graphics files
  \DeclareGraphicsExtensions{.eps}     % for pure latex we expect eps files
\fi%

\graphicspath{{figures/}{pictures/}{images/}{./}} % where to search for the images

\usepackage{microtype}                 % use micro-typography (slightly more compact, better to read)
\PassOptionsToPackage{warn}{textcomp}  % to address font issues with \textrightarrow
\usepackage{textcomp}                  % use better special symbols
\usepackage{mathptmx}                  % use matching math font
\usepackage{times}                     % we use Times as the main font
\usepackage{cite}                      % needed to automatically sort the references
\usepackage{tabu}                      % only used for the table example
\usepackage{booktabs}                  % only used for the table example
\usepackage{enumitem}

\newcommand{\dhk}[1] 
{\textcolor{black}{#1}}

\newcommand{\modified}[1] 
{\textcolor{black}{#1}}
\onlineid{1900}

\vgtccategory{Research}

\vgtcinsertpkg

\title{Comparative Analysis of Change Blindness in Virtual Reality and Augmented Reality Environments}
\author{Donghoon Kim\thanks{e-mail: donghoon.kim@usu.edu}\\ %
        \scriptsize Utah State University %
\and Dongyun Han\thanks{e-mail: dongyun.han@usu.edu}\\ %
     \scriptsize Utah State University %
\and Isaac Cho\thanks{e-mail: isaac.cho@usu.edu}\\ %
     {\scriptsize Utah State University }}

\abstract{
Change blindness is a phenomenon where an individual fails to notice alterations in a visual scene when a change occurs during a brief interruption or distraction.
Understanding this phenomenon is specifically important for the technique that uses a visual stimulus, such as Virtual Reality (VR) or Augmented Reality (AR).
Previous research had primarily focused on 2D environments or conducted limited controlled experiments in 3D immersive environments.
In this paper, we design and conduct two formal user experiments to investigate the effects of different visual attention-disrupting conditions (Flickering and Head-Turning) and object alternative conditions (Removal, Color Alteration, \modified{and Size Alteration}) on change blindness detection in VR and AR environments.
Our results reveal that participants detected changes more quickly and had a higher detection rate with Flickering compared to Head-Turning. Furthermore, they spent less time detecting changes when an object disappeared compared to changes in color or size. Additionally, we provide a comparison of the results between VR and AR environments. 

} % end of abstract

\CCScatlist{
  \CCScatTwelve{Human-centered computing}{Visu\-al\-iza\-tion}{Visu\-al\-iza\-tion techniques}{Treemaps};
  \CCScatTwelve{Human-centered computing}{Visu\-al\-iza\-tion}{Visualization design and evaluation methods}{}
}

\begin{document}

%% The ``\maketitle'' command must be the first command after the
%% ``\begin{document}'' command. It prepares and prints the title block.

%% the only exception to this rule is the \firstsection command

\maketitle

% Definition
\section{Introduction}
% \HDY{I think the paragraphs are unnecessarily separated. try to make this section have three or four paragaphs. maybe + 1 more paragraphs summarizing what our contributions are }

% [Change Detection in Virtual Reality]
% Visual stimulus
% Change Detection
% Change Blindness
% What can affect to Change Blindness?
A visual stimulus, a virtual environment created by computer graphics, is the primary component of immersive technologies \cite{cummings2016immersive}.
% It delivers visual information to provide an immersive experience to the user in a virtual environment in Virtual Reality (VR) and Augmented Reality (AR).
It delivers visual information to provide an immersive experience to a Virtual Reality (VR) and Augmented Reality (AR) user. 
% These technologies are subject to the same limitations as the human visual system encountered in the real world.
The user in these immersive environments is subject to have the same limitations as the human visual system encountered in the real world.
% such as the minimum time to perceive the visual stimulus\cite{potter2014detecting}. 
% Change blindness~\cite{potter2014detecting} serves as an example of such limitations. 
%The phenomenon of change blindness, in particular, exemplifies how such limitations can significantly impact human life.
% [About Change blindness]
% \ic{add clear motivation about this research}
%Change blindness
% Change blindness, as a phenomenon exemplifying such limitations, refers to the failure of individuals to notice significant changes in their environments \cite{potter2014detecting}. 
Change blindness, for example, refers to the failure of individuals to notice significant changes in an environment \cite{potter2014detecting}.
These changes include object displacement or removal from the Field of View (FoV), alterations in the appearances of observers, or complete scene transformations. 
%situations in which an object is moved or removed from view, an observer's appearance is altered, or an entire scene is transformed.
% This phenomenon could occur in various situations, such as when an object is moved or removed from sight, when an observer's appearance is altered, or when an entire scene is transformed.
Change blindness can even occur when the changes are directly in front of individuals \cite{rensink1997see,simons1997change}.

% [Why it is important]
% Detecting what is changed, how much has or has not been changed, is one of the most necessary capacities to understand a situation as visual information.
% Most people guess that they can figure out the changed things; however, the change blindness phenomenon makes it wrong.
% Particularly in time-sensitive, irreversible contexts, such as vehicular navigation, criminal investigations, or military operations \cite{romer2014adolescence, white2010blind, davies2007change, nelson2011change, fitzgerald2016change, divita2004verification}, change blindness can pose substantial challenges, and its reduction is crucial.
Change blindness presents substantial challenges, particularly in time-sensitive and irreversible scenarios like vehicular navigation, criminal investigations, or military operations \cite{romer2014adolescence, white2010blind, davies2007change, nelson2011change, fitzgerald2016change, divita2004verification}.
% Despite the change detection that overcomes change blindness is possible through training, because it is limited to the specific area, the systematic handling for change blindness is necessary for visual content
While the earlier studies show that training can temporarily improve change detection such as pop-out detection \cite{ahissar1996learning}, 
its effectiveness is limited to specific domains, and errors can still occur \cite{gaspar2013change}. Therefore, it is crucial to adopt a systematic approach to fully understand change blindness itself and improve visual content.

The emergence of immersive technologies, especially Head-Mounted Displays (HMDs) has opened a new era of visual perception research. It enables unlimited alteration and control of variables in immersive environments. Virtual environments facilitate numerous alterations that would be unattainable in the real world, allowing for the investigation of change blindness under various conditions. Employing object alterations, such as modifying object orientation, removing objects, or changing  a room structure, may be unfeasible in a real-world setting. It can be readily used within VR environments\cite{steinicke2011change, charlton2013driving, suma2011leveraging}.
% \ic{add research questions we are trying to attach}
%However, the current VR/AR HMDs have several limitations. Numerous investigations, theories, and experiments have been conducted to comprehend the change blindness phenomenon. However, the quantitative relationship between the intensity of change blindness and the object manipulation method, or the strategy employed to disrupt visual attention, remains ambiguous.
However, VR/AR HMDs have limited FoV~\cite{baudisch2003halo, piumsomboon2017covar}.
The FoV refers to the extent of the visual environment that a user can perceive through the HMDs.
Their limited FoV could restrict the user's ability to perceive changes in their surroundings if the changes occur outside the FoV.
Moreover, the quantitative relationship between the change blindness effects and the object alteration conditions or attention-disrupting conditions remains unclear.
% \ic{this may good to have a image to show different FoVs of those HMDs compared to humans' FoV}

\begin{figure*}[t]
\centering
\includegraphics[width=.95\textwidth]{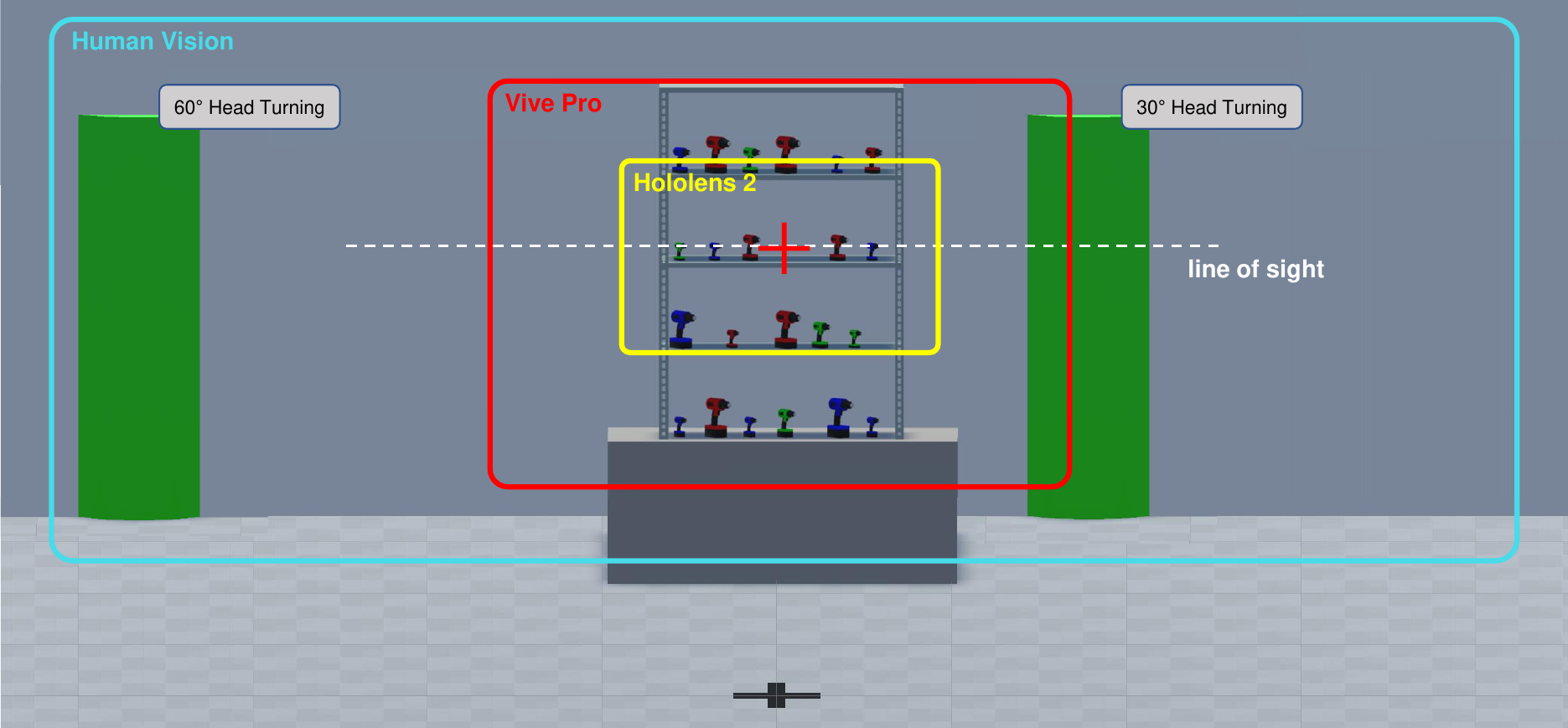}
\caption{
Effective FoV of the Vive Pro, Hololens 2, and human vision.
The FoV in the apparatuses' spec catalog is not consistent with experience.
The rectangles represent a visual boundary that an observer can see through the HMDs.
The height of the shelf is modified following the height of the participant to their line of sight is the same with medium size objects' center on the third floor.
Two columns on the left and right are located symmetrically on the position for 60$^\circ$ and 30$^\circ$ Head-Turning conditions.
In the Head-Turning conditions, a participant has to turn their head until one of the two columns (left or right) is located center of their sight.
The default color of cylinders is white (right side in Figure), but when the participant's center of view (the red cross in the center of the figure) reaches it, the object is altered, and the color is changed to green (left side in Figure) to notice the object's alternation.
A black cross line on the bottom is a participant's standing position.
}
\label{figure:FoV}   
% \vspace{-0.3cm}
\end{figure*}

To address these research gaps, we establish the following research questions (RQs):
%Addressing the ambiguous aforementioned, we establish the following research questions (RQs):

\begin{itemize}[leftmargin=*, noitemsep]
    % \item \textbf{RQ1}. Does the change blindness phenomenon exist in a factitious virtual environment?

    \item \textbf{RQ1:} How do the limited FoVs influence change detection in immersive environments, and how does the influence differ between VR and AR?

    \item \textbf{RQ2:} How do different types of object alteration methods, such as color change, size modification, or displacement, affect change detection in both VR and AR environments?
%    \item \textbf{RQ3}.What are the differences in effects of change blindness between VR and AR?  % About VR and AR comparison 

\end{itemize}

% [What we will discuss]
To tackle the research questions, we designed and conducted two user studies. In Study 1, we evaluated different types of visual attention-disrupting methods (Flickering and Head-Turning) and object manipulation methods (Object Removal, Color Alteration, \modified{and Size Alteration}) in VR. Study 2 evaluated the same conditions with an additional Head-Turning angle in AR.% Our results shed light on the impact of independent variables on change blindness and highlight differences between VR and AR. 

Our results show %the change blindness phenomenon in VR and AR environments and the effects of each condition.
that participants identified changes faster when an object was altered within FoV than when the object was altered outside of FoV. Change detection was more challenging when the object attributes, such as color or size were altered, as opposed to when the object was entirely removed. %It appears that change blindness can readily occur when the alternation was on out of FoV.
%Additionally, the type of alternation was one of the major parts of the change blindness phenomenon. 
Additionally, the results show differences in change detection between AR and VR environments, highlighting the distinct effects of different immersive technologies.

\section{Related Work}
\subsection{Change Blindness}

Change blindness is a visual perception phenomenon that occurs when a stimulus changes without being noticed by observers. Earlier psychology and cognitive studies reported change blindness is caused due to the limitation of visual short-term memory~\cite{brady2011review, ungerleider1998neural, bays2009precision, castillo2020allocentric, pertzov2012forgetting}. According to a coherence theory and triadic architecture~\cite{rensink2002change, rensink2000dynamic}, a human visual attention involves three stages including early processing, focused attention, and release of attention. The early processing stage is a low-level processing that takes place in parallel across the entire visual environment. It generates proto-objects which is with weak spatial and temporal coherence. In the focused attention stage, the generated proto-objects are then selected and combined into spatial and temporal individual object images. These images enable humans to understand the continuity of objects across interruptions. Finally, they are dissolved when attention is released. Detecting changes in visual stimuli is highly relevant to the focused attention stage.

Change blindness occurs when the focused attention stage is interrupted~\cite{simons2005change, levin1997failure, rensink2000visual, rensink2002failure}.
Earlier research investigated what could divert focused attention and result in change blindness.
Flickering is a widely used attention-disrupting method in change blindness studies in 2D display settings~\cite{rensink1997see}.
It distracts observers by repeatedly displaying a blank image between the original and altered images.
%(Fig~\ref{}).
Changing an object while the entire scene is moving~\cite{suchow2011motion} and asking observers to perform complex tasks~\cite{simons1999gorillas, triesch2003you} also could confuse the observers in deciding where to focus.
Moreover, if an object is changed outside of the observers' sight, the observers may not have a chance to pay attention.
Some research investigated change blindness in VR, and their details are reported in the next subsection.
%Martin et al.~\cite{martin2023study} demonstrated that it could cause more intense change blindness.
% \HDY{In this work, we investigate the effect of flikering and changing outside of observers' sight on change blindness in VR and AR environments because …}

\subsection{Change Blindness in Virtual Environments}
Previous research reported that change blindness could also occur in virtual environments.
Steinicke et al.\cite{steinicke2011change} investigated change blindness phenomena in various stereoscopic display systems including Active Stereoscopic Workbench, Passive Back-Projection Wall, and VR HMD device using flickering as a distraction method. They reported that change blindness has occurred in all types of display systems, and their participants have different response times to find changes depending on the displays. Triesch et al.~\cite{triesch2003you} investigated change blindness in a virtual environment asking participants to move virtual objects or change their sizes. Their findings showed that as task complexity increases, so does change blindness.

Change blindness in VR is investigated as a method to render a larger virtual environment in a limited physical space~\cite{suma2011leveraging}.
Suma et al.~\cite{suma2010exploiting} made participants recognize virtual space larger than the actual tracking area utilizing change blindness.
The authors changed the position of a door when the participant was exploring a room to make them go out in a different direction than entered.
Consequently, while the participants are visiting many different rooms, they only walk roundly in a limited tracking space and perceive the virtual area as larger than the real one.
Surprisingly, the participants failed to detect these changes, resulting in an illusion in which they perceived themselves to be in a vast virtual environment.

Factors that can either amplify or diminish change blindness remain underexplored.
The majority of change blindness research has only used manipulation methods that add or remove objects in a scene.
It would be because of the limitations of the study environments.
% For example, changing the color or size of an object, such as the color of a tree or car and the size of a building or animal, without any other difference in a short time is almost impossible in the real world.
For example, changing the color or size of an object, such as a tree, car, building, or animal, in a short time is not applicable in the real world.
However, change blindness can be explored further in VR beyond such limitations.
Martin et al.~\cite{martin2023study} recently investigated the effect of object shapes, colors, and locations on change blindness.
% They report that the factors (altered object's complexity, distance between the object and observer, alteration type, and alteration timing) have an impact on change blindness.
They experimented with factors (altered object's complexity, distance between the object and observer, alteration type, and alteration timing) that can affect change blindness, and they found the effects of distance between the object and observer and the complexity of the altered object for the change detection ratio.
However, they could not find an effect of alteration types which is found in the change blindness studies in 2 dimensional~\cite{rensink1997see, ma2013change}.
In our work, we investigated what visual attention-disrupting and object-alteration conditions can reinforce or discourage the change blindness phenomenon in the VR and AR environment with identical shape objects on shelves as grid-like to free from any other uncontrolled variables.

\begin{figure*}[t]
\centering
\includegraphics[width=0.9\textwidth]{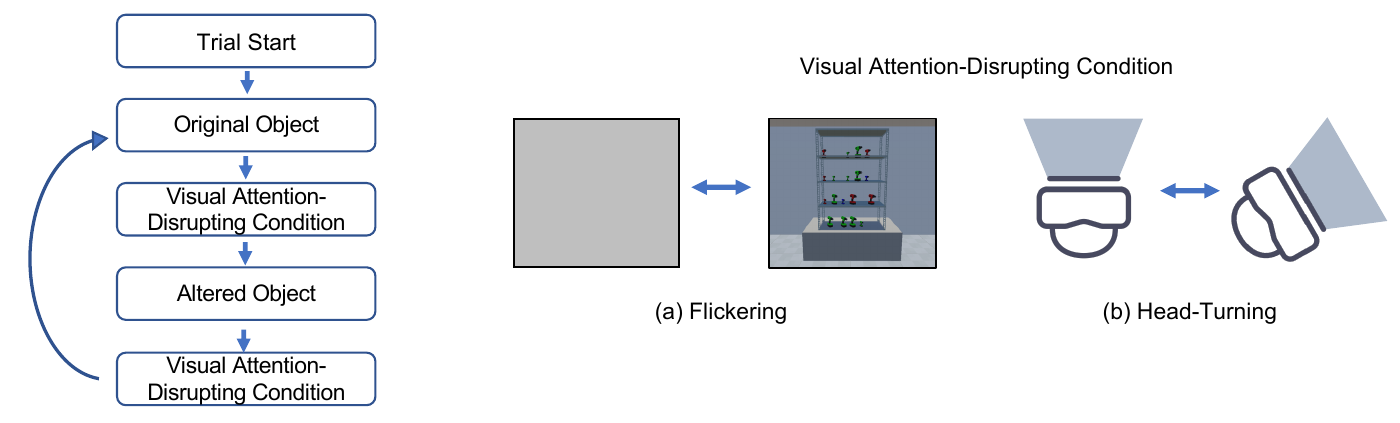}
\caption{
The experimental procedures involve presenting the original object among other static objects at the beginning of the task. During the task, participants experience disruptions in their visual attention either in the flickering condition (a) or the head-turning condition (b), the object undergoes alteration based on the designated object alternation condition. Through the task, participants sequentially see the original object and the altered object until they make a selection indicating which object they believe has been altered.
% Participants can select the object they guess it is changed anytime in the task.
} 
\label{Picture_Front_Flicker}   
% \vspace{-0.3cm}
\end{figure*}

% \begin{figure*}[t]
% \centering
% \includegraphics[width=\textwidth]{Figure/Final/Alternations.pdf}
% \caption{Object's variations: Large, Medium, Small sizes; Red, Blue, Green colors} 
% \label{Picture_Objects}   
% \end{figure*}

\begin{figure}[t]
\centering
\includegraphics[width=\columnwidth]{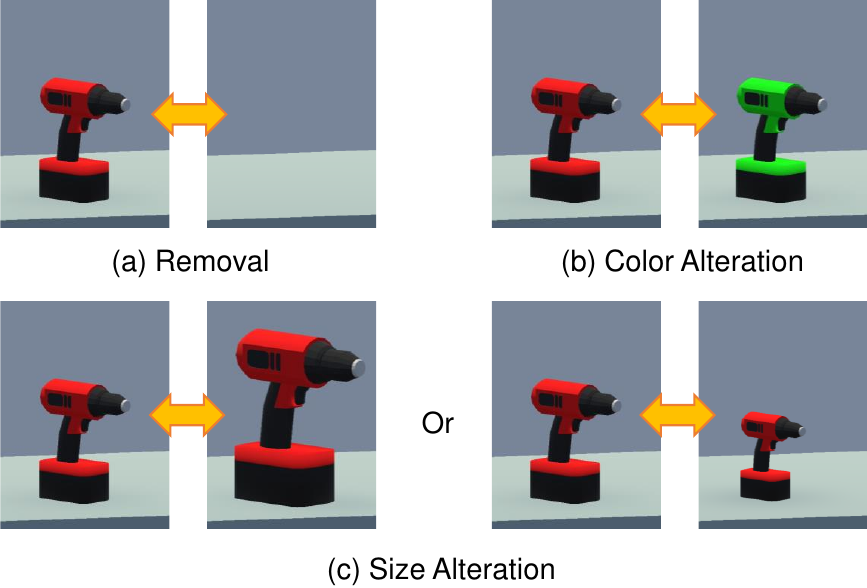}
\caption{Object alternation conditions: a) Object Removal - the object is removed and only shows the background; b) Color Alteration - the object's color is changed to one of other colors (red, green, or blue). The changed color is fixed during a task; c) \modified{Size Alteration - the object size is altered between medium to large (left) or medium to small (right).}} 
\label{Picture_Objects}   
% \vspace{-.3cm}
\end{figure}

\section{Experiment Design}
The primary objective of our research is to investigate the impact of limited FoV of VR/AR HMDs and object alteration methods on change blindness in VR/AR environments.
To achieve this goal, we conduct two controlled experiments.
%user studies.

In the first experiment, we evaluate two visual attention-disrupting conditions (Flickering and 60$^\circ$ Head-Turning) and four object alternation conditions (Object Removal, Color Alteration, \modified{and Size Alteration}) in VR. 
The second experiment investigates the impact of the same conditions with an additional visual attention-disrupting condition (30$^\circ$ Head-Turning) in an AR environment.

Previous change blindness studies in a VR environment reported unclear results with different shapes of manipulated objects on different positions and not uniformed background images\cite{martin2023study}.
Therefore, we design an experiment scene to investigate the change blindness phenomenon with identical shape objects (hand drills) and unobtrusive background design (monochromatic gray wall).

%\ic{AR}
%*VR environment has virtual objects only -> But the AR system can provide virtual objects in the real world -> Change blindness study in AR can have more real-world like study (or how the real-world environment can disrupt the user attention on the virtual object or limited FOV of AR device? <- If the number of objects are larger than the human limitation (6~7), it can be meaningless.)

%This study aimed to examine the effect of real-world background interference by using an experimental environment that was kept the same as Study 1, except for the virtual room.
% This experiment was conducted in front of a large gray screen that was similar to the back wall in the virtual environment in Study 1.

% \subsection{Experiment Design}

\subsection{Experiment Environment}
The virtual scene is a 10m $\times$ 10m room with a shelf (Figure~\ref{figure:FoV}). In VR, the scene has a monochromatic gray background, while there is no background in AR. 
The shelf has four layers, each containing multiple drill objects that can be altered.
These drills can be in three different sizes \modified{(small: 3cm (width) $\times$ 6cm (height) $\times$ 10cm (depth), medium: 5cm $\times$ 10cm $\times$ 15cm, or large: 7cm $\times$ 14cm $\times$ 20cm)} and colors (red, green, or blue).
Participants can select one drill object during a trial that they believe has been altered.
To prevent participants from anticipating the changes, the position, size, and color of the drills are randomized.
An altered object's color and size are also randomized for each trial to make the altered object unpredictable for the participants.
For the Head-turning condition, two gray columns are located on the left and right sides of the shelf and change their color to green to notice participants that the object is altered when they turn their head. 

%In the AR environment, the virtual objects are the same composition as the VR environment except for the virtual walls, floor, and ceiling.
%The room was a 4m x 4m and a 3.4m x 1.8m gray background screen is hanging on the wall in which the participant will head.

The distance between a participant and the shelf is 2.5m, and the position of the shelf is set as the center of objects on the 3rd shelf to match it with the participant's eye level (Figure~\ref{figure:distance_and_info_panel}).

% Reviewer 5's Comment 4: 4) Related to 3): The shelf is positioned in front of the observer. Have you considered changes in size in the periphery? I.e., the objects may subtend different angles from the user's point of view/
%\modified{
%The width of the shelf is 1.5m, much smaller than the distance between this and the participant; thus, we ignore the horizontal displacement.}
An information panel is located between the participant and the shelf to show the  visual attention-disrupting condition and the remaining number of tasks.
The participant starts a trial by clicking a  button located on the bottom of the panel. 

%cipant has interaction (aim and press a button on the VR controller (VR) or pinch gesture (AR)), a task will start and the panel is disappeared, and it comes back after each task end.

%The position, size (small, medium, large), and color (of to be changed object are randomly changed on each task to avoid anticipation from participants.

\subsection{Visual Attention-Disrupting Conditions}
To evaluate the impact of the limited FoV of VR/AR HMDs on change blindness, we use two visual attention-disrupting conditions: 

\begin{itemize}[ leftmargin=*, noitemsep]
\item \textbf{Flickering:} It is a traditional method used to investigate change blindness \cite{rensink1997see, simons2005change}.
In this condition, a monochromatic gray barrier periodically obstructs an observer's view, and an object is altered while the observer's sight is blocked.
The barrier is designed to block the observer's sight for 250 ms, and this blocking period repeats every 500 ms.
This flickering time is chosen with the task difficulty and the range of human eye blinking in mind, which typically falls between 100 ms and 400 ms~\cite{beanland2017change, martin2023study}.
% This flickering time is selected to consider the task difficulty and the range of human eye blinking which typically falls between 100ms to 400ms \cite{beanland2017change, martin2023study}.

\item \textbf{Head-Turning:} In this condition, an observer is required to voluntarily rotate his or her head to the left or right until the shelf is outside the FoV and an object is altered when the shelf is out of FoV.
In the VR environment, a Head-Turning angle of $60^\circ$ to ensure that the shelf is completely outside of the FoV ($110^\circ$).
Similarly, due to the narrower FoV ($50^\circ$) in the AR environment, \modified{a reduced $30^\circ$ Head-Turning angle is also used to ensure that the alteration is occurred outside of the FoV.}
\modified{Additionally, to ensure clarity of the alteration that occurred outside the FoV, the turning angle is quantified based on the horizontal angle of the HMD, and the observer is required to stand stationary in a location.}
However, a Head-turning angle of $60 ^\circ$ is also used to enable a direct comparison between VR and AR conditions.
To ensure that the observer is aware that he or she has turned his or her head sufficiently, the virtual scene includes columns on the left and right sides that change color from red to green when the turning angle exceeds the set angle (Figure~\ref{figure:FoV}).
This visual feedback serves as a confirmation that the required Head-Turning motion has been performed adequately.

%Once the shelf is not visible, and the turning angle reaches a set angle (60\textdegree for the VR experiment and 60\textdegree and 30\textdegree for the AR experiment), the object alteration takes place.
\end{itemize}

\subsection{Object alternation Conditions}

Object alteration occurs when participants' visual attention is disrupted as one of these two visual attention-disrupting conditions.
One of the drills on the shelf will be changed by one of the following four alternation methods: 

\begin{itemize}[leftmargin=*, noitemsep]
\item \textbf{Object Removal:} 
The target object temporarily vanishes from the scene and then reappears.
% This condition temporarily removes the target object from the scene while disrupting the observer's visual attention and reintroduces it.
\item \textbf{Color Alteration:}
This condition changes the target object's color to a different one (i.e., red, blue, or green) from its original color.
% while the observer's visual attention is disrupted. 
% \item \textbf{Size Incrementation} 
% The target object's size is increased from small to medium or medium to large (Fig.~\ref{Picture_Objects}).
% This condition enhances the target object's size incrementally from small to medium or from medium to large.
% \item \textbf{Size Decrementation}
% This condition diminishes the target object's size from large to medium or from medium to small.
% This condition diminishes the target object's size incrementally from large to medium or from medium to small.
\item \textbf{Size Alteration:} 
\modified{This condition involves altering the target object's size, either transitioning between small and medium or medium and large.}

\end{itemize}

In the VR experiment, each participant is asked to complete a total of 40 \modified{tasks (5 trials each for Removal and Color Alteration conditions + 10 trials for Size Alteration condition) $\times$ 2 visual attention-disrupting conditions) to find and select an altered object.}
Similarly, in the AR experiment, each participant is asked to complete a total of 60 tasks \modified{(5 trials each for Removal and Color Alteration conditions + 10 trials for Size Alteration condition) $\times$ 3 visual attention-disrupting conditions).}

\modified{It is worth noting that we initially conducted the study with two types of Size Alteration conditions: medium to small, and medium to large. Upon analysis, however, both were found to be equivalent. As a result,  we merged the two conditions resulting in a doubled size of trials for the Size Alteration condition.}

\subsection{Measurements}
% \ic{need a paragraph to introduce overall measures, something like }
We employ multiple measures to evaluate the impact of limited FoV and object alteration conditions.
These measures offer a comprehensive understanding of participants' performance and their subject experience through the experiments.
% In this study, the system records the time between the beginning of each task and the moment of the participant's notice.
% During the tasks, the participant's visual attention is recorded by the eye tracking technique, and the number manipulation of the object is recorded.

% Change blindness refers to the inability of individuals to detect significant changes in visual scenes, even when the changes are substantial.

\textbf{Detection Time}: 
It is the duration from the start of a task until the participant identifies and selects the altered object.
This measurement assesses the time it takes for the participant to perceive and recognize alteration in each task.
% This is a measure used to assess the time it takes for participants to perceive and recognize changes in visual scenes in an immersive environment.
A shorter detection time indicates a quicker and more efficient perception and recognition of alteration. 
% A shorter detection time indicates a quicker and more efficient perception and recognition of changes suggesting a reduced presence of change blindness.
On the other hand, a longer detection time suggests a delayed perception and recognition of alteration, indicating a higher presence of change blindness.
%Please note that
If a participant fails to detect the altered object within the designated time limit (60 seconds), that case is excluded from the detection time analysis. 
%If participants can detect changes quickly, it represents a weaker presence of change blindness, while a longer response time indicates a more substantial presence of change blindness.

\textbf{Detection Rate}:
In a trial, a participant's selection may not always be correct. 
% During the trials, the participant may select an object that they believe has been altered, but his or her selection may not always be accurate. 
The detection rate assesses the participant's ability to correctly identify the altered objects within the given time limit.
% for each condition. 
It is a ratio of the number of correct selections to the total number of responses, except for the detection failure cases.
% It is a ratio between the number of responses within the time limit that correspond to the correct selection and the total number of responses made by the participant.
A higher detection rate indicates that the participant could remember the objects accurately without misconceiving them.
% A higher detection rate indicates that the participant has a higher level of accuracy in identifying the altered object, reflecting a strong ability to detect changes in the virtual scene.
%This measures understanding how well the participants can find without confusion.

\textbf{Timeout Rate}:
If a participant fails to detect the altered object within the time limit, we consider this as the occurrence of change blindness.
The timeout rate is the ratio of the number of timeouts to the total number of trials for each condition.
A higher timeout rate indicates a higher occurrence of change blindness. 
%We set the time limit as 60 seconds, and this measurement shows the ratio of the timeout tasks per the total task for each condition.

% \textbf{gazing time - flickering}: 
% One interesting characteristic of change blindness is that even though the observer sees the changed object, they can not easily figure it out.
% Measuring how long time the participants gazed at the changed object per each task can find this case.

\textbf{Head-Turning Count}:
This measurement counts how many times a participant turns his or her head during the trials in the Head-Turning condition. In the Flickering condition, the alternation count is proportional to the response time. In the Head-Turning condition, the Head-Turning count can indicate the alternation count.
% Thus, the number of the object changing is proportional to the response time.
% \dhk{Unlike with the flickering condition, which automatically disrupts sight, the object-alternation}
% But with the turning method, a participant can handle when the target object changes.
% Consequently, this measurement, the number of head-turning, can be found how many times the target object is changed until the participant selects the altered object.

\textbf{Participants' preference and feedback}: 
% After finishing every set of tasks (20 tasks with a single visual attention-disruption method) and all tasks (40 tasks for the VR experiment and 60 tasks for the AR experiment)
After finishing all the tasks, the participant is asked to do questionnaires.
The questionnaires ask which manipulation method was the hardest or the easiest, which visual attention-disruption method was the hardest, and what strategy they used to detect the change.

\subsection{Procedures}
% Upon arrival, a participant is presented with an informed consent form and asked to read and sign the form.
Upon arrival, a participant is asked to read and sign an informed consent form according to the IRB protocol (IRB VR: \#12970, AR: \#13243).
The participant then completes a demographic questionnaire. 
% Following that, they are a brief explanation about the experiment's purpose and procedures as well as the alteration conditions, and visual attention-disrupting conditions.
Following that, the participant receives a description of the experiment's purpose and procedures, visual attention-disrupting conditions, and object alteration conditions.
A training session in an immersive VR or AR environment is provided to the participant.
In the training session, the participant is instructed on how to select an object using either a controller with a ray casting (VR experiment) or his or her hand with a pinch gesture (AR experiment).
During the training session, the participant familiarizes with the visual attention-disrupting conditions. The participant is not required to physically move around the scenes during the experiments. 
% The participant is not required physically move around during the trials.
%The task is to identify and select the altered object, and in the head-turning condition, the participant needs to turn his or her head accordingly. 

\begin{figure}[t]
\centering
\includegraphics[width=\columnwidth]{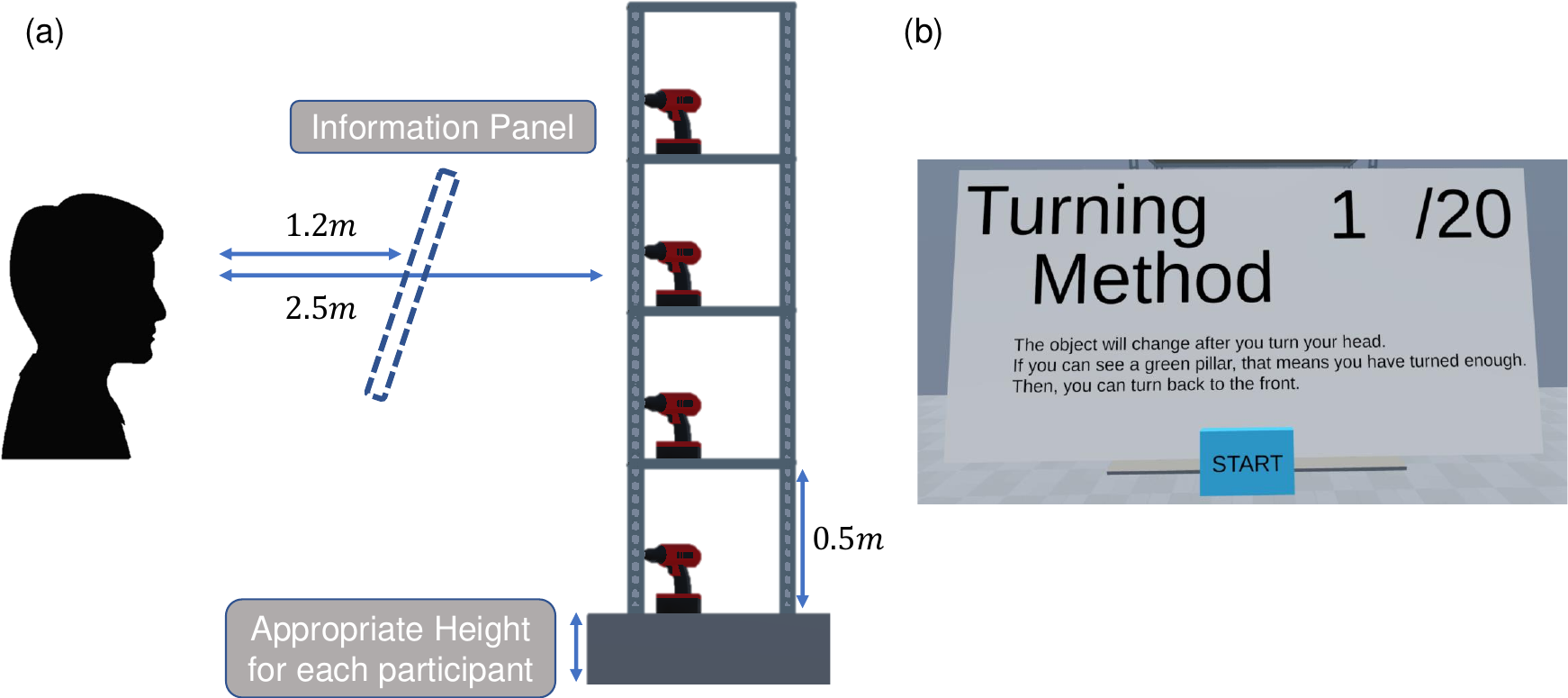}
\caption{
(a) Distance between the participant and the shelf is 2.5m, and while in the experiment, a participant does not get out of the initial position.
The information panel is located center between them; the elevation of the shelf is controlled to match the participant's eye level and the center of objects on the 3rd shelf.
(b) Information panel provides details about what visual attention-disrupting condition will be used and how many tasks are left.
% \ic{change the text size of the figure to be consistent with other figures}
} 
\label{figure:distance_and_info_panel}
% \vspace{-0.3cm}
\end{figure}

The main study is conducted after the training session.
%, consisting of 40 (VR experiment) or 60 (AR experiment) trials (2 or 3 attention-disrupting methods \texttimes 4 object changing methods \texttimes 5 trials). 
For each visual attention-disrupting condition, the participant is required to complete 20 trials for each object alteration condition.
% The participant has only 20 trials at once with one visual attention-disrupting method.
%The order of the object alternation conditions is fully randomized.
A new trial begins when the participant selects a start button on the information panel (Figure~\ref{figure:distance_and_info_panel}).
The information panel shows the current visual attention-disrupting condition and the number of remaining trials.
%However, what manipulation method is used is not provided.
For each trial, 16 $\sim$ 21 drill objects are randomly generated on the shelf with random positions, colors, and sizes. The participant is asked to find an altered object as fast and correctly as possible. 
% The participant's task is to find the object that has been altered while the visual attention is disrupted by flickering or head turning as fast and correctly as possible. 

%In the flickering condition, a grayboard blocks the participant's sight 250ms, and participants can search for the object 250ms between the blocking (The right side of Figure~\ref{Picture_Front_Flicker}.)

%Under the turning method, there is no grayboard; the object is changed when the participant turns their head. When they have turned enough angle (60\textdegree or 30\textdegree depends on the experiment condition), the green sign notice they turned enough (Figure~\ref{Picture_Turning}.) The participant can turn their head on which side they want (left or right).

After completing all trials with a specific visual attention-disrupting condition, the participant is asked to complete a questionnaire.
The questionnaire asks about the experience of the visual attention-disrupting condition, object alteration condition, and the strategy for identifying the altered object during the trials.
Next, the participant undergoes the same procedure with another visual attention-disrupting condition.
%To achieve counterbalance, the order of the visual attention-disrupting method uses all permutations of the methods ($_{2}P_{1}$ for the VR experiment and $_{3}P_{1}$ for the AR experiment).
After all the trials, the participant is asked to complete a post-questionnaire to compare the difficulty of visual attention-disrupting methods and their reason. The order of the visual attention disruption and object alteration conditions is fully counterbalanced by the Latin square.

% In the AR environment, the participant has to use their hand for interaction instead of the controller. It makes hard the far distance interaction because when the user tries to have gestures for interaction, the aim is shaken.
% To solve this problem, study 2 uses a gaze laser from the center of the forehead. The user has to aim using that gaze laser and interact using finger gestures.

\subsection{Hypothesis}
% Based on the experiment design, 
We set the following five hypotheses.

\begin{enumerate} [label=\textbf{H\arabic*:}, noitemsep]
% \item In VR and AR, Head-Turning would have a longer detection time, less detection rate, and higher timeout rate than Flickering.
% This is because it takes more time for participants to physically turn their heads and refocus on the scene after head movement.
\item In both VR and AR, Head-Turning would have a longer detection time, lower detection rate, and higher timeout rate than Flickering because participants will take more time to physically turn their heads and refocus on the scene after head movement.
% The head-turning condition is expected to result in a longer detection time compared to the flickering condition in both VR and AR. 
% This is due to the fact that additional time is required for participants to physically turn their heads and to refocus on the scene after head movement.
% Furthermore, the head-turning condition necessitates participants to refocus on the scene after the head movement.
% \item In both VR and AR environments, the removal condition is expected to result in a shorter detection time and higher detection rate compared to other alteration conditions.
\item In both VR and AR environments, the Removal condition is expected to result in a shorter detection time and higher detection rate compared to other alteration conditions.
\dhk{
This is because the object's spatial information can be perceived non-attentionally, but other visual channels are not, according to triadic architecture~\cite{rensink2000dynamic}.
% We guess that this non-attentional perception can make change detection efficient with the removal condition.
}
% The object removal alternation condition is expected to result in a shorter detection time and higher detection rate compared to other alteration conditions. Because, as we can perceive the fast-passing object was passed but cannot notice its shape or color, the object's existence is the primary attribute in visual perception.
\item In both VR and AR, each \dhk{visual channel} of an object, such as color, size, or shape, is expected to have different perception difficulties and make a different result of change detection.
\dhk{This is because visual channels, size, and color have traditionally been regarded as separate in visual perception\cite{ware2019information, smart2019measuring}.}

% This is because participants require more time to rotate their head and refocus engaging in larger angle head-turning, as compared to smaller angle head-turning.
%\item The centered-located object's alternation is expected to result in a shorter response time. This is due to the boundary of our sight having lower perceiving strength than the center.

\item In the AR environment, Head-Turning with a larger angle has a longer detection time, a lower detection rate, and a higher timeout rate than Head-Turning with a smaller angle.
We assume the large turning angle makes refocusing on the scene harder than the smaller angle.

\item Participants have longer detection times, lower detection rates, and higher timeout rates in AR than in VR. \dhk{Because despite being identical virtual objects, the participant's perception of them differs between VR and AR\cite{jones2008effects, gaffary2017ar}.
}

% the AR environment can distract the user's attention from the virtual object's transparency and a real-virtual mixed environment
% In addition, the previous study was in the VR environment and was only conducted with virtual objects.
% However, despite being identical virtual objects, the observer's perception of them differs between VR and AR environments.\cite{jones2008effects, gaffary2017ar}
% \item (VR/AR) The change detection is dependent on the observer's finding strategy, such as where to find first, and how much focus on it.

% AR is expected to disrupt change detection compared to VR. Besides the FoV problem, Virtual objects in AR have limitations, such as transparency and a real-virtual mixed environment, making it harder for object perception than in VR.
% AR is expected to disrupt change detection compared to VR. Besides the FoV problem, Virtual objects in AR have limitations, such as transparency and a real-virtual mixed environment, making it harder for object perception than in VR.

% \item A smaller FoV environment is expected to obstruct unexpected change detection compared to a large FoV environment. The small FoV makes the participants attend to a specific point, but the large FoV allows them to have a wider visual search.
\end{enumerate}
%Head turning
%

\subsection{Apparatus}

In the VR experiment, the Vive Pro HMD with a wireless adapter and a single controller is utilized.
It features a 110$^\circ$ vertical and horizontal FoV and its resolution is 1440 x 1600 pixels per eye (2880 x 1600 pixels combined).
The wireless adapter offers near-zero latency.
In the AR experiment, Microsoft HoloLens 2 is used.
It features a 43$^\circ$ horizontal and 29$^\circ$ vertical FoV, a resolution of 2048 × 1080 for each eye, and a refresh rate of 75Hz.
% It also supports hand-tracking technology, allowing user input through hand gestures.
According to Sauer et al.~\cite{sauer2022assessment}, however, these manufacturer FoV angles are not consistent with real-world use.
% However, these FoV angles that the manufacturer claimed are not consistent with the real user experiment\cite{sauer2022assessment}.
The true FoV boundaries of the HMDs are shown in Fig~\ref{figure:FoV}.
% The real FoV boundary is presenting on Fig~\ref{figure:FoV}.
The application for both experiments is executed in Unity 2021.3.7f1 and ran on a Windows 10 desktop with Intel Xeon W-2245 CPU (3.90GHz), 64GB RAM, and Nvidia GeForce RTX 3090 graphics card.
% \modified{The average luminosity of tracking space is about 250 lux, and the brightness control setting on the Hololens 2 device is 100\%.}
\modified{The average luminosity of tracking space is about 250 lux, and the brightness setting on Hololens 2 is 100\%.}
\section{Results}
% \ic{add a summary paragraph here about this section}\\
% \HDY{unify the terms! ex, change all 'object-manipulation' to 'object alteration'}

% To analyze the data collected from our user studies,
In this section, we report user study results.
We use a two-way (visual attention-disrupting x object alternation) repeated measures Analysis of Variance (ANOVA) test at a significance level of 5\% for the detection time, detection rate, and timeout rate analyses. A
one-way repeated measures ANOVA test at the same significance level of 5\% is used for Head-Turning count and qualitative analyses.

In this experiment, we aim to investigate the change blindness phenomenon in a VR environment, which is affected by object manipulation conditions and visual attention-disrupting conditions.
Previous studies about change blindness in a VR environment reported unclear results with different shapes of alternated objects on different positions and not uniformed background images\cite{martin2023study}.
Therefore, in this experiment, we designed a VR experiment scene to investigate the change blindness phenomenon with the same shape objects and unobtrusive background design.
\subsection{Experiment 1: VR}
\subsubsection{Participants}
A total of 22 participants (11 males and 11 females) are recruited from the university's participant recruitment system (SONA). %Minimun # by G power is 24
Their average age is 19.7, ranged from 18 to 22.
All participants have 20/20 (or corrected 20/20) vision, and they do not have impairments in using VR devices. %, such as color blindness.
% We reward 1 SONA credit to each participant as follows the university SONA policy. 
They were rewarded 1 SONA credit as follows the university SONA policy. According to the pre-questionnaire, 18 out of 22 participants have experience using the VR device.
% , and their self-evaluated VR familiarity score is 2.95 out of 7.0.
% According to the pre-questionnaire, 18 out of 22 participants (82\%) have experience using a VR device, and their self-evaluated VR familiarity score is 2.95 out of 7.0.

\subsubsection{Quantitative Analysis}

%We use a two-way repeated measures Analysis of Variance (ANOVA) test for response time and accuracy analysis and a one-way repeated measures ANOVA test for turning number analysis at a significance level of 5\%.
%\textcolor{red}{Huynh-Feldt's correction is applied to degrees of freedom to account for violations of the spherical assumption.} % if the data does not have the spherical assumption. 
%We use Fisher's Least Significant Difference (LSD) for post-hoc comparisons at the same significant level. 

%---------------------------------------- Response Time ---------------------------------
%---------------------------------------- Response Time - Sight x Manipulation ---------------------------------

\begin{figure}[t]
\centering
\includegraphics[angle=270, width=.47\textwidth]{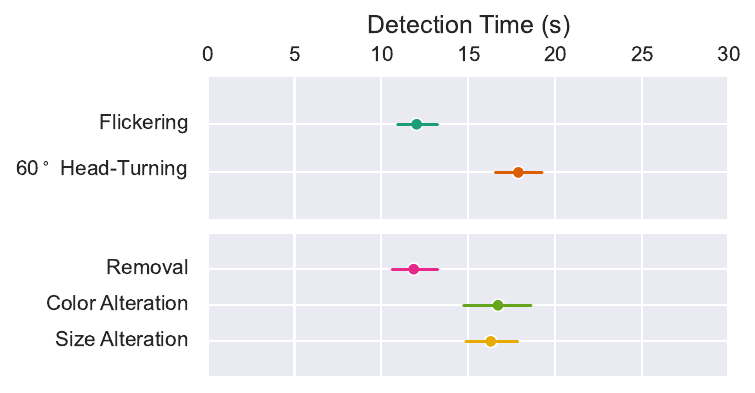}
\caption{
VR Experiment: Detection Time Results (95\% CI error bar)
% with the two visual attention-disrupting conditions (Upper) and the four object alternation conditions (Lower).
% The detection time with the flickering condition is significantly faster than the 60$^\circ$ head-turning condition.
% And the result of the removal condition is also significantly faster than the other three alternation conditions.
}
\label{figure:vr_responseTime_sight_manipulation}   
% \vspace{-0.4cm}
\end{figure}

%---------------------------------------- Detection Time Table ---------------------------------

% \begin{table}[ht] %Data with "VR_Response_Time.pdf"
% \caption{VR Experiment: Statistical detection time results with the visual attention-disrupting (V) and object alternation (A) conditions.
% }
% \begin{tabular} {m{2.3cm} | m{1.1cm} | m{1cm}| m{0.8cm} | m{0.8cm}}
% \toprule 
%     {\textbf{Factor}} & {\textbf{$DOF$}}  & {\textbf{$F$}} & {\textbf{$p$}} & {\textbf{$\eta$\textsubscript{$p$}\textsuperscript{$2$}}}\\ \midrule
%     Interaction (V\texttimes A)  & {3, 63} & {0.345}  & {.793} & {.016} \\ 
%     V  & {1, 21} &  {31.763}  & {\textless .001} & {.602} \\
%     A & {3, 63}  & {12.604} & {\textless .001} & {.375} \\
%     \bottomrule
% \end{tabular}
% \label{table:vr_responseTime_sight_manipulation} 
% \end{table}

%---------------------------------------- Statistical result Table ---------------------------------

\textbf{Detection Time:} 
% The statistical results of response time with the visual attention-disrupting \ic{condition} and the object \ic{alteration condition} are reported in Table~\ref{table:vr_responseTime_sight_manipulation}.
\dhk{
The results with each condition are reported in Figure~\ref{figure:vr_responseTime_sight_manipulation}.% and its statistical results are reported in Table~\ref{table:vr_responseTime_sight_manipulation}.
}
No interaction effect is disclosed between the visual attention-disrupting conditions and the object alternation conditions \modified{($p$=.873)}.
We find a main effect on the visual attention-disrupting conditions \modified{(F(1,21)=32.7, $p$\textless .001, \textbf{$\eta$\textsubscript{$p$}\textsuperscript{$2$}}=.609).
Flickering has a faster detection time (M=12.0s, SD=4.90) than Head-Turning (M=17.9s, SD=5.58)}. 
In addition, a main effect on the object alternation conditions \modified{(F(2,42)=18.7, p $<$ .001, \textbf{$\eta$\textsubscript{$p$}\textsuperscript{$2$}}=.471) is also found.} 
Its pairwise comparison results show that Removal \modified{(M=11.9s, SD=4.58) has a faster detection time result than the result with Color Alteration (M=16.7s, SD=6.56, p\textless .001) and Size Alternation (M=16.3s, SD=5.48, p\textless .001).}

\begin{figure}[t]
\centering
\includegraphics[angle=270, width=.47\textwidth]{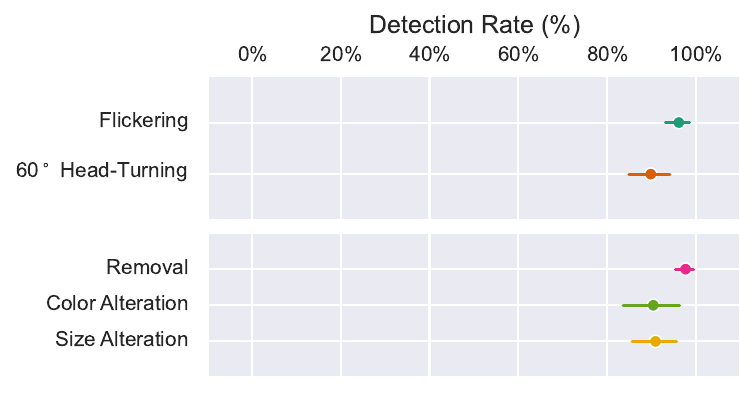}
\caption{
VR Experiment: Detection Rate Results (95\% CI error bar)
% with the two visual attention-disrupting conditions (Upper) and the four object alternation conditions (Lower).
% The detection rate with the flickering condition is significantly more accurate than the 60$^\circ$ head-turning condition.
% And the result of the removal condition is also significantly more accurate than the other three alternation conditions.
} 
\label{figure:vr_accuracy_sight_manipulation}   
\end{figure}

%---------------------------------------- Statistical result Table ---------------------------------

% \begin{table}[ht]
% \caption{VR Experiment: Statistical detection rate results with the visual attention-disrupting (V) and object alternation (A) conditions.
% }
% \begin{tabular} {m{2.3cm} | m{1.1cm} | m{1cm}| m{0.8cm} | m{0.8cm}}
% \toprule 
%     {\textbf{Factor}} & {\textbf{$DOF$}}  & {\textbf{$F$}} & {\textbf{$p$}} & {\textbf{$\eta$\textsubscript{$p$}\textsuperscript{$2$}}}\\ \midrule
%     Interaction(V\texttimes A) & {3, 63} & {2.919}  & {.041} & {.122} \\ 
%     V  & {1, 21} &  {5.121}  & {.034} & {.196} \\
%     A & {3, 63}  & {3.971} & {.012} & {.159} \\
%     \bottomrule
% \end{tabular}
% \label{table:vr_accuracy_sight_manipulation}
% %\vspace{-0.6cm}
% \end{table}

\textbf{Detection Rate:}
% The detection rate results with each condition are reported in Figure~\ref{figure:vr_accuracy_sight_manipulation}.
Figure~\ref{figure:vr_accuracy_sight_manipulation} shows the detection rate results with each condition.
% \modified{\sout{We find an interaction effect on the detection rate between the visual attention-disrupting and the object alteration conditions (F(2,42)=5.179, p=.010, \textbf{$\eta$\textsubscript{$p$}\textsuperscript{$2$}}=.198).
% In the Head-Turning condition, the participants have a better detection rate with the Removal (M=96.4\%, SD=9.79 (9.79)) condition than the Size Alternation (M=97.5\%, SD=X.X, $p$ \textless .001) conditions.
% However, in the Flickering condition, we found no statistical differences among object alteration conditions.
% In the Size Alteration conditions, the Flickering condition (M=99.1\%, SD=4.17 (Incrementation), M=96.4\%, SD=7.71 (Decrementation)) has a higher detection rate than the 60$^\circ$ Head-Turning (M=84.5\%, SD=22.5, $p$=.006 (Incrementation), M=83.6\%, SD=25.9, $p$=.034 (Decrementation)).}}
Simple effects on the object alternation conditions (F(2, 42)=5.90, $p$ = .006, \textbf{$\eta$\textsubscript{$p$}\textsuperscript{$2$}}=.219) are disclosed.
% \modified{\sout{About the visual attention-disrupting conditions, the Flickering condition (M=96.6\%, SD=11.0) made a higher detection rate result than the 60$^\circ$ Head-Turning condition (M=88.4\%, SD=21.6, $p$=.034).}}
Regarding the object alternation conditions, Removal (M=97.7\%, SD=7.65) produced a higher detection rate result than the results with Color Alteration (M=90.5\%, SD=20.7, $p$=.017) \modified{and Size Alteration (M=91.0\%, SD=17.0, $p$=.001).}

%----------------------------------- Accuracy - Position ---------------------------------
% In Figure~\ref{figure:vr_accuracy_position_sight}, we find that the participants' detection rate does not relate to the object locations (i.e., level and place), unlike the detection time result.

% \begin{figure}[t]
% \centering
% \includegraphics[angle=270, width=.47\textwidth]{Figure/Final/VR_Accuracy_Position.pdf}
% \caption{
% VR Experiment: Detection rate results with the altered object locations.
% } 
% \label{figure:vr_accuracy_position_sight}   
% \end{figure}

% \begin{table}[ht] 
% \caption{\label{table:vr_accuracy_position} \textbf{VR Experiment: Statistical accuracy result with the manipulated object position.}}
% \begin{tabular} {m{2.3cm} | m{1.4cm} | m{1cm}| m{0.8cm} | m{0.8cm}}
% \toprule 
%     {\textbf{Factor}} & {\textbf{$DOF$}}  & {\textbf{$F$}} & {\textbf{$p$}} & {\textbf{$\eta$\textsubscript{$p$}\textsuperscript{$2$}}}\\ \midrule
%     Floor  & {3, 63} &  {1.432}  & {.242} & {.064} \\
%     Side & {11, 165}  & {.514} & {.892} & {.033} \\
%     \bottomrule
% \end{tabular}
% \end{table}

% The statistical accuracy results with the object position are reported in Table~\ref{table:vr_accuracy_position}.

%----------------------------------- Timeout ---------------------------------
%----------------------------------- Timeout - Sight x Manipulation ---------------------------------
\begin{figure}[t]
\centering
\includegraphics[angle=270, width=.47\textwidth]{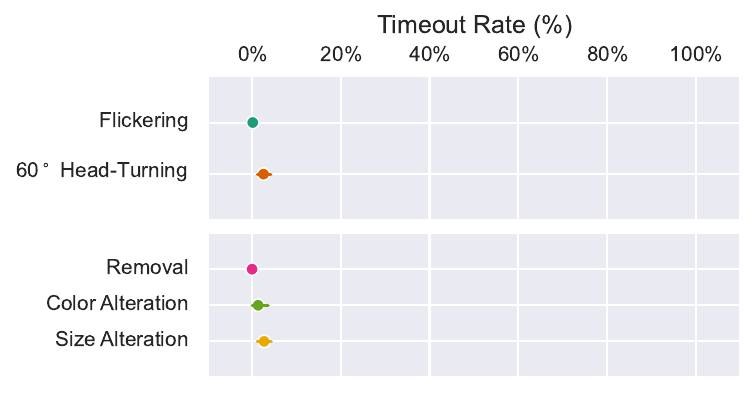}
\caption{
VR Experiment: Timeout Rate Result (95\% CI error bar)
% with the visual attention-disrupting and object alternation conditions.
% The timeout rate with the flickering condition is significantly less than the 60$^\circ$ head-turning condition.
} 
\label{figure:vr_timeout_sight_manipulation}   
\vspace{-0.4cm}
\end{figure}

%---------------------------------------- Statistical result Table ---------------------------------

% \begin{table}[ht] %
% \caption{VR Experiment: Statistic timeout rate results with the visual attention-disrupting and the object alternation conditions.}
% \begin{tabular} {m{2.3cm} | m{1.1cm} | m{1cm}| m{0.8cm} | m{0.8cm}}
% \toprule 
%     {\textbf{Factor}} & {\textbf{$DOF$}}  & {\textbf{$F$}} & {\textbf{$p$}} & {\textbf{$\eta$\textsubscript{$p$}\textsuperscript{$2$}}}\\ \midrule
%     Interaction (V\texttimes A)  & {3, 63} & {2.135}  & {.105} & {.092} \\ 
%     V  & {1, 21} &  {8.351}  & {.009} & {.285} \\
%     A & {3, 63}  & {2.545} & {.064} & {.108} \\
%     \bottomrule
% \end{tabular}
% \label{table:vr_timeout_sight_manipulation} 
% %\vspace{-0.6cm}
% \end{table}

\textbf{Timeout Rate:}
% The statistical response time results of the participant who could not find and select object case ratios (timeout ratio) are reported in Table~\ref{table:vr_timeout_sight_manipulation}.
\dhk{
The timeout rate results with each condition are reported in Figure~\ref{figure:vr_timeout_sight_manipulation}.% and its statistical results are reported in Table~\ref{table:vr_timeout_sight_manipulation}.
}
% The statistical results are reported in Table~\ref{table:vr_timeout_sight_manipulation}.
\modified{These results show an interaction effect between the visual attention-disrupting and the object alteration conditions (F(2,42)=3.22, $p$=.050, \textbf{$\eta$\textsubscript{$p$}\textsuperscript{$2$}}=.133).
In the Head-Turning condition, the participants have a lower timeout rate with the Removal (M=0.00\%, SD=0.00) condition than the Size Alteration (M=5.00\%, SD=6.57 $p$=.002) condition.
With the Size Alteration condition, the Flickering (M=0.50\%, SD=2.08) condition has a lower timeout rate than the Head-Turning ($p$=.005) %(M=5.00\%, SD=6.57, $p$=.005) 
condition.
% simple effect on head-turing condition and size alteration condition 
 There is a simple effect on the visual attention-disrupting condition (F(1,21)=6.37, $p$=.020, \textbf{$\eta$\textsubscript{$p$}\textsuperscript{$2$}}=.233). The Flickering condition shows a lower timeout rate (M=0.152\%, SD=1.22) than the Head-Turning condition (2.58\%, SD=6.81). There is another simple effect on the object alteration (F(2,42)=4.11, $p$=.023, \textbf{$\eta$\textsubscript{$p$}\textsuperscript{$2$}}=.164) conditions. The removal condition has a lower timeout rate (M=0.00\%, SD=0.00) than the Size Alteration condition (M=2.73\%, SD=5.38).}

\begin{figure}[t]
\centering
\includegraphics[angle=270, width=.47\textwidth]{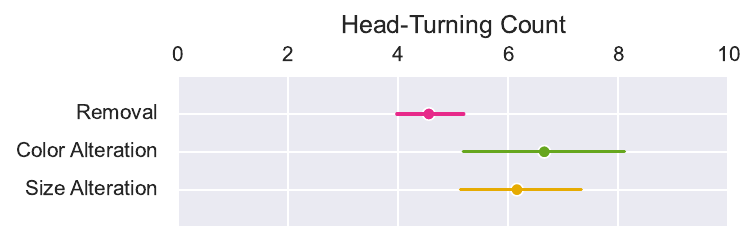}
\caption{VR Experiment: Head-turning Count Result (95\% CI error bar)
% with the object alternation conditions.
} 
\label{figure:vr_turning_manipulation}   
% \vspace{-0.4cm}
\end{figure}

%---------------------------------------- Statistical result Table ---------------------------------

% \begin{table}[ht] %
% \caption{
% VR Experiment: Statistic head-turning count result with the object alternation (A) conditions.
% }
% \begin{tabular} {m{2.3cm} | m{1.1cm} | m{1cm}| m{0.8cm} | m{0.8cm}}
% \toprule 
%     {\textbf{Factor}} & {\textbf{$DOF$}}  & {\textbf{$F$}} & {\textbf{$p$}} & {\textbf{$\eta$\textsubscript{$p$}\textsuperscript{$2$}}}\\ \midrule
%     A  & {3, 63} &  {5.939}  & {.001} & {.220} \\
%     \bottomrule
% \end{tabular}
% \label{table:vr_turning_manipulation}
% %\vspace{-0.6cm}
% \end{table}

\textbf{Head-Turning Count:} 
% The statistical head-turning count with manipulation methods result is reported in Table~\ref{table:vr_turning_manipulation}.
The results are reported in Figure~\ref{figure:vr_turning_manipulation}.
% The head-turning count results with each condition are reported in Figure~\ref{figure:vr_turning_manipulation}.
% We report the statistical results in Table~\ref{table:vr_turning_manipulation}.
\modified{
A main effect is found on the object alternation conditions (F(2, 42)=8.67, p\textless.001, \textbf{$\eta$\textsubscript{$p$}\textsuperscript{$2$}}=.292); the participants made less number of Head-Turning with Removal (M=4.56, SD=1.47) than Color Alteration (M=6.66, SD=3.42 $p$=.003) and Size Alteration (M=6.16, SD=2.49, $p$\textless.001).
}
\subsubsection{Qualitative Analysis}
% We asked questionnaires about the experience after the end of each visual attention-disrupting condition.
% The participants responded the hardest visual attention-disrupting condition is the head-turning condition (19 participants answered).
We find a significant difference in the subjective difficulty in finding the altered object (F(1,42) = 4.776, $p$ = .034, \textbf{$\eta$\textsubscript{$p$}\textsuperscript{$2$}} = .102) between Flickering (M = 3.27) and Head-Turning (M = 4.27) from 1 (easiest) to 7 (hardest).
Many participants commented that they could easily remember the objects as images in their heads within the Flickering condition.
But within the Head-Turning condition, they noted it was not easy to remember and to keep their eyes focused on the objects. P15 commented \textit{``When I turned my head, it was a lot harder to pay attention to the changes because I would slowly forget the details of the environment.''}. And P22 stated \textit{``Self turning was more difficult. The change was less noticeable and harder to recognize whether that be color or size. With the faster flickering, I was able to quickly spot the change whether that be size or color.  I could focus on rows at a time with this method or sometimes the whole picture and notice it. However with self turning, I had to focus on smaller parts of the picture and focus on either size or color change, not both.''}

Among the object alteration conditions, the participants rated the Removal condition as the easiest condition, 16 participants chose it for the Flickering condition, while 14 selected it for the Head-Turning condition. \modified{Conversely, the Size Alteration condition was consistently regarded as the most challenging, with 16 participants selecting it under both the Flickering and Head-Turning conditions.}
%participants within flickering; 16 (incrementation: 6, decrementation: 10) participants within head-turning).

% And mostly, 
%Most of the participants responded with a similar search strategy: searching from top to bottom and trying to remember row by row. Here we share some participants' comments on this VR experiment. % related to the search strategy and conditions.
\begin{quote}
% \textit{P10 - Strategy: ``At first, I was trying to look at the whole shelf at once. However, I found it to be much easier when I focused on one shelf at a time and memorized the colors (even though I was mainly focusing on colors, I was still able to notice when the size changed), then move on to the next shelf if I couldn't see anything.''}

% \textit{P17 - Strategy: ``I would start at the top shelf and quickly memorize what I saw. Then, once I turned my head, if anything changed I would click on it. If not I would go to the next row. ''}

% \textit{P15: ``When I turned my head, it was a lot harder to pay attention to the changes because I would slowly forget the details of the environment.''}

% \textit{P22: ``Self turning was more difficult. The change was less noticeable and harder to recognize whether that be color or size. With the faster flickering, I was able to quickly spot the change whether that be size or color.  I could focus on rows at a time with this method or sometimes the whole picture and notice it. However with self turning, I had to focus on smaller parts of the picture and focus on either size or color change, not both.''}

\end{quote}

\subsection{Experiment 2: AR}

\subsubsection{Participants}
A total of 22 participants (10 males and 12 females) are recruited from SONA. %Minimun # by G power is 24
Their average age is 20, ranging from 18 to 24.
All participants have 20/20 (or corrected 20/20) vision, and they do not have impairments in using AR devices.%, such as color blindness.
We reward 2 SONA credits to each participant as follows the university SONA policy.
According to the pre-questionnaire, 10 out of 22 participants (45.5\%) have experience using an AR device.
\modified {None of the participants from the VR study took part in the AR study.}
%, and their self-evaluated AR familiarity score is 2.93 out of 7.0.

\subsubsection{Quantitative Analysis}

%We use a two-way repeated measures ANOVA test for response time and accuracy analysis same as the VR experiment.
%A one-way repeated measures ANOVA test is used for the head-turning count analysis, and both tests are at a significance level of 5\%.

% ---------------------AR - Response Time-------------------
% ---------------------AR - Response Time - Sight x Manipulation------------------

\begin{figure}[t]
\centering
\includegraphics[angle=270, width=.47\textwidth]{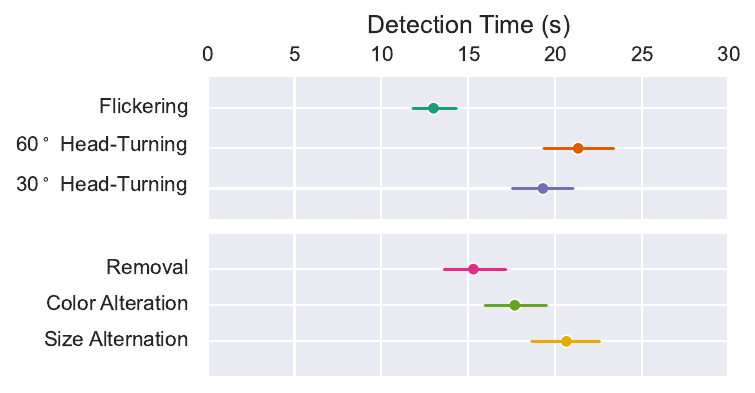}
\caption{AR Experiment: Detection Time Results (95\% CI error bar)
% with the three visual attention-disrupting conditions (Upper) and the four object alternation conditions (Lower).
% The detection time result with the flickering condition is significantly faster than the results with both head-turning conditions. And the result of the removal condition is also significantly faster than the other three object alternation conditions.
} 
\label{figure:ar_responseTime_sight_manipulation}   
% \vspace{-0.3cm}
\end{figure}

%---------------------------------------- Statistical result Table ---------------------------------

% \begin{table}[ht]
% \caption{AR Experiment: Statistical detection time results with the visual attention-disrupting (V) and object alternation (A) conditions.}
% \begin{tabular} {m{2.3cm} | m{1.1cm} | m{1cm}| m{0.8cm} | m{0.8cm}}
% \toprule 
%     {\textbf{Factor}} & {\textbf{$DOF$}}  & {\textbf{$F$}} & {\textbf{$p$}} & {\textbf{$\eta$\textsubscript{$p$}\textsuperscript{$2$}}}\\ \midrule
%     Interaction(V \texttimes A)  & {6, 126} & {1.923}  & {.082} & {.084} \\ 
%     V  & {2, 42} &  {19.395}  & {\textless .001} & {.480} \\
%     A & {3, 63}  & {16.921} & {\textless .001} & {.446} \\
%     \bottomrule
% \end{tabular}
% \label{table:ar_responseTime_sight_manipulation} 
% %\vspace{-0.6cm}
% \end{table}

\textbf{Detection Time:}
\dhk{
The detection time results with each condition are reported in Figure~\ref{figure:ar_responseTime_sight_manipulation}.% and its statistical results are in Table~\ref{table:ar_responseTime_sight_manipulation}.
}
% Table~\ref{table:ar_responseTime_sight_manipulation} and Figure~\ref{fig:ar_responseTime_sight_manipulation} show the results.
% The statistical results of response time with the AR experiment are reported in Table~\ref{table:ar_responseTime_sight_manipulation}, and Figure~\ref{fig:ar_responseTime_sight_manipulation} shows those results for each visual attention-disrupting method and object-manipulation method.
No significant interaction effect is disclosed.
% Even though the p-value for the interaction is not a significant effect ($p$=.082), it is significantly smaller than the p-value for the interaction in the VR study ($p$=.793).

\modified{
Two main effects from the visual attention-disrupting conditions (F(2,42)=18.1, $p$\textless .001, \textbf{$\eta$\textsubscript{$p$}\textsuperscript{$2$}}=.463) and object alternation conditions (F(2,42)=21.6, $p$\textless .001, \textbf{$\eta$\textsubscript{$p$}\textsuperscript{$2$}}=.504) are found. Flickering has a faster detection time (M=13.0s, SD=5.15) than 30$^\circ$ Head-Turning (M=19.3s, SD=7.63, $p$\textless .001), and 60$^\circ$ Head-Turning (M=21.3s, SD=8.29, $p$\textless .001).
Removal has a faster detection time (M=15.3s, SD=7.77) than Color Alteration (M=17.7s, SD=7.38, $p$=.005) and Size Alteration (M=20.7s, SD=7.86, $p$\textless .001).
In addition, Color Alteration (M=17.7s, SD=7.38) has a faster detection time than Size Alteration (M=20.7s, SD=7.86, $p$=.001).
}

\begin{figure}[t]
\centering
\includegraphics[angle=270, width=.47\textwidth]{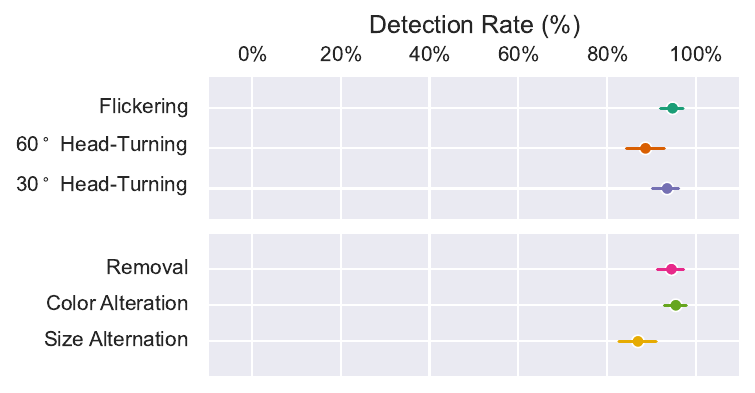}
\caption{
AR Experiment: Detection Rate Results (95\% CI error bar)
% with the three visual attention-disrupting conditions (Upper) and the four object alternation conditions (Lower).
% The results with the visual attention-disrupting conditions do not present a statistically significant difference.
% But the result of the removal condition is significantly higher than the results with the size alternation conditions.
% And the result of the color alteration condition is also significantly higher than those.
% Average response time result (Left) and Accuracy (Right) for two sight-blocking methods and four object-object changing methods. The response times, which is the time the participant needs to notice the changed object for each combination of conditions, and the accuracy, which is the accuracy of the participant's response, show significantly different results depending on the conditions.
} 
\label{figure:ar_accuracy_sight_manipulation}   
% \vspace{-0.4cm}
\end{figure}

%---------------------------------------- Statistical result Table ---------------------------------

% \begin{table}[ht]
% \caption{
% AR Experiment: Statistical detection rate results with the visual attention-disrupting (V) and object alternation (A) conditions.
% }
% \begin{tabular} {m{2.3cm} | m{1.1cm} | m{1cm}| m{0.8cm} | m{0.8cm}}
% \toprule 
%     {\textbf{Factor}} & {\textbf{$DOF$}}  & {\textbf{$F$}} & {\textbf{$p$}} & {\textbf{$\eta$\textsubscript{$p$}\textsuperscript{$2$}}}\\ \midrule
%     Interaction(V\texttimes A)  & {6, 126} & {2.067}  & {.062} & {.090} \\ 
%     V  & {2, 42} &  {3.131}  & {.054} & {.130} \\
%     A & {3, 63}  & {5.867} & {.001} & {.218} \\
%     \bottomrule
% \end{tabular}
% \label{table:ar_accuracy_sight_manipulation}
% %\vspace{-0.6cm}
% \end{table}

\textbf{Detection Rate:}
\dhk{The detection rate results are reported in Figure~\ref{figure:ar_accuracy_sight_manipulation}. % and its statistical results are reported in Table~\ref{table:ar_accuracy_sight_manipulation}.
}
% We report the statistical results in Table~\ref{table:ar_accuracy_sight_manipulation}, and Figure~\ref{fig:ar_responseTime_sight_manipulation}.
% The statistical results of accuracy with the visual attention-disrupting methods and the object-manipulation methods are reported in Table~\ref{table:ar_accuracy_sight_manipulation}, and Figure~\ref{fig:ar_responseTime_sight_manipulation} presents its results.
\modified{
The results only disclose the main effect of the object-alternation conditions (F(2, 42)=7.15, $p$=.002, \textbf{$\eta$\textsubscript{$p$}\textsuperscript{$2$}}=.254).
Pairwise comparisons show that Removal (M=94.5\%, SD=12.2) has a higher detection rate than Size Alteration (M=87.0\%, SD=18.6, $p$=.011). Color Alteration (M=95.5\%, SD=10.6) also has a higher detection rate than Size Alteration (M=87.0\%, SD=18.6, $p$\textless.001).
}
\begin{figure}[t]
\centering
\includegraphics[angle=270, width=.47\textwidth]{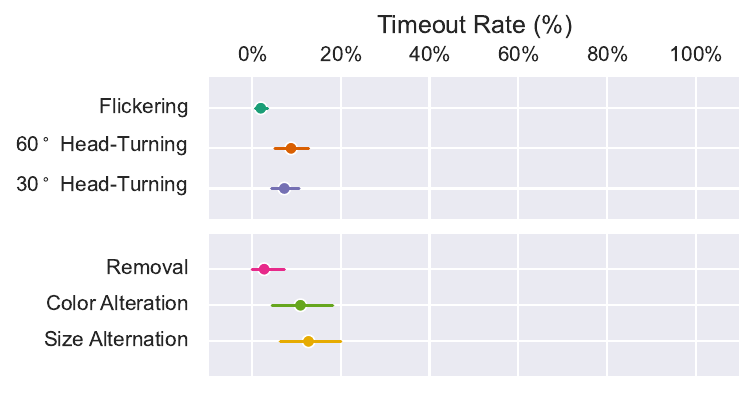}
\caption{
AR Experiment: Timeout Rate Results (95\% CI error bar)
% with the visual attention-disrupting conditions and object alternation conditions.
% The timeout rate result with the flickering condition is significantly less rate than the results with the 60$^\circ$ and 30$^\circ$ head-turning conditions.
% The result with the removal condition is also significantly less rate than results with other object alternation conditions.
} 
\label{figure:ar_timeout_sight_manipulation}   
% \vspace{-0.3cm}
\end{figure}

%---------------------------------------- Statistical result Table ---------------------------------

% \begin{table}[ht] %
% \caption{
% AR Experiment: Statistical timeout results with the visual attention disrupting (V) and the object-alternation (A) conditions.
% }
% \begin{tabular} {m{2.3cm} | m{1.1cm} | m{1cm}| m{0.8cm} | m{0.8cm}}
% \toprule 
%     {\textbf{Factor}} & {\textbf{$DOF$}}  & {\textbf{$F$}} & {\textbf{$p$}} & {\textbf{$\eta$\textsubscript{$p$}\textsuperscript{$2$}}}\\ \midrule
%     Interaction (V\texttimes A)  & {6, 126} & {1.917}  & {.083} & {.084} \\ 
%     V  & {2, 42} &  {4.159}  & {.022} & {.165} \\
%     A & {3, 63}  & {7.021} & {\textless .001} & {.251} \\
%     \bottomrule
% \end{tabular}
% \label{table:ar_timeout_sight_manipulation}
% %\vspace{-0.6cm}
% \end{table}

\textbf{Timeout Rate:}
\dhk{
Figure~\ref{figure:ar_timeout_sight_manipulation} presents the timeout rate results with each condition.
}
% with the visual attention-disrupting methods and the object-manipulation methods.
\modified{
Two main effects are found on the visual attention-disrupting conditions (F(2, 42)=4.03, $p$=.025, \textbf{$\eta$\textsubscript{$p$}\textsuperscript{$2$}}=.161) and the object-alteration conditions (F(2, 42)=9.90, $p$\textless .001, \textbf{$\eta$\textsubscript{$p$}\textsuperscript{$2$}}=.320).
Flickering (M=1.97\%, SD=5.57) has less timeout rate than both Head-Turning conditions (30$^\circ$: M=7.30\%, SD=12.7, \textbf{$\eta$\textsubscript{$p$}\textsuperscript{$2$}}=.014; 60$^\circ$: M=8.79\%, SD=15.4, \textbf{$\eta$\textsubscript{$p$}\textsuperscript{$2$}}=.006).
We also find Removal (M=1.52\%, SD=6.33) has a significantly lower timeout rate than Color Alteration (M=7.27\%, SD=12.9, $p$=.006) and Size Alteration (M=9.24\%, SD=14.8, $p$\textless .001).
}
% And when the object is manipulated with the removal method (M=1.52\%, SD=6.33), it has fewer timeout cases than the color (M=7.27\%, SD=12.9, $p$=.006), size-increase (M=9.70\%, SD=15.7, $p$\textless .001), and size-decrease (M=8.79\%, SD=16.7, $p$=.003) methods.
% ------------------ Timeout - Position ------------------
% Similar to the accuracy result (Table~\ref{table:ar_accuracy_position}), 

% No significant effects of the altered object location are observed on the timeout rate results.

% \begin{figure}[t]
% \centering
% \includegraphics[angle=270, width=.47\textwidth]{Figure/Final/AR_TimeOut_Sight_Position.pdf}
% \caption{AR Experiment: Timeout rate result with the altered object locations.
% } 
% \label{fig:ar_timeout_sight_position}   
% % \vspace{-0.1cm}
% \end{figure}

% However, in the pairwise comparison, the removal manipulation method shows a significantly shorter gazing time than the size manipulation methods (Increase: 1.05\%, $p$=.044; Decrease: 1.30\%, $p$=.014)

% ---------------------AR - Head-turning count -------------------
% ---------------------AR - Head-turning count - Manipulation ------------------

\begin{figure}[t]
\centering
\includegraphics[angle=270, width=.47\textwidth]{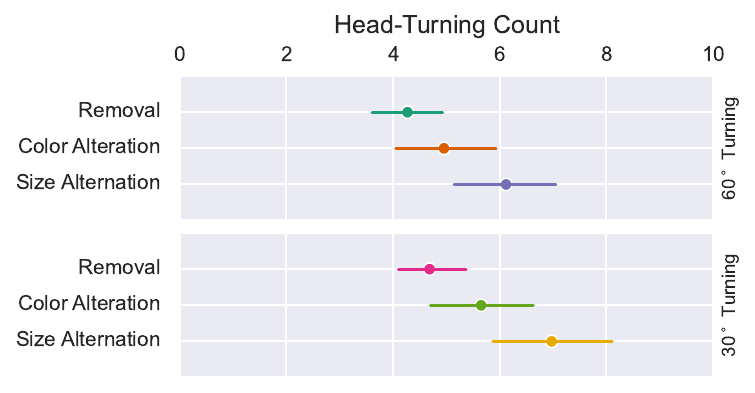}
\caption{
AR Experiment: Head-Turning Count Result (95\% CI error bar)
% with the object-manipulation methods.
} 
\label{figure:ar_turning_sight_manipulation}   
% \vspace{-0.4cm}
\end{figure}

%---------------------------------------- Statistical result Table ---------------------------------

% \begin{table}[ht]
% \caption{
% AR Experiment: Statistic head-turning count results with the angle of the turning methods and the object manipulation method.
% }
% \begin{tabular} {m{2.5cm} | m{1.1cm} | m{1cm}| m{0.8cm} | m{0.8cm}}
% \toprule 
%     {\textbf{Factor}} & {\textbf{$DOF$}}  & {\textbf{$F$}} & {\textbf{$p$}} & {\textbf{$\eta$\textsubscript{$p$}\textsuperscript{$2$}}}\\ \midrule
%     Angle\texttimes Manipulation  & {3, 63} & {3.394}  & {.023} & {.139} \\ 
%     Angle  & {1, 21} &  {3.065}  & {.095} & {.127} \\
%     Manipulation & {3, 63}  & {9.743} & {\textless .001} & {.317} \\
%     \bottomrule
% \end{tabular}
% \label{table:ar_turning_sight_manipulation}
% %\vspace{-0.6cm}
% \end{table}

\textbf{Head-Turning Count:}
% \modified{\sout{The Head-Turning and the object-alteration conditions show an interaction effect (F(3, 63)=3.394, $p$=.023, \textbf{$\eta$\textsubscript{$p$}\textsuperscript{$2$}}=.129).
% With Turning-60$^\circ$, Removal (M=4.27, SD=1.58) has less Head-Turning than Size Incrementation (M=5.53, SD=2.54,$p$=.020) and Decrementation (M=6.71, SD=2.43, $p$ < .001).
% The Color Alteration (M=4.96, SD=2.38) and Size Incrementation conditions also have less Head-Tuning count than Size Decrementation, showing $p$=.004 and $p$=.017 respectively. 
% With Turning-30$^\circ$, similar to Turning-60$^\circ$, Removal (M=4.69, SD=1.52) has less Head-Turning count than Size Incrementation (M=7.45, SD=3.97, $p$=.005), Size Decrementation (M=6.46, SD=2.36, $p$=.007) and Color Alteration (M=5.65, SD=2.35, $p$=.048).
% In addition, Color Alteration has fewer Head-Turning counts than Size Incrementation ($p$=.012), but no difference with Size Decrementation.
% In the Size Incrementation condition, Turning-60$^\circ$ (M=5.53, SD=2.54) has less Head-Turning than Turning-30$^\circ$ ($p$=.018).}}
\modified{
A main effect is found on the object-alteration condition (F(2, 42)=12.4, $p$\textless .001, \textbf{$\eta$\textsubscript{$p$}\textsuperscript{$2$}}=.371). 
Removal (M=4.48, SD=1.56, $p$\textless.001) and Color Alteration (M=5.31, SD=2.39, $p$=.002) made less Head-Turning than Size Alteration (M=6.55, SD=2.54). 
% Color Alteration (M=5.305, SD=2.388) also has less Head-Turning than Size Alteration (M=6.548, SD=2.535, $p$=.002).
% Removal (M=4.479, SD=1.563) made less Head-Turning than Size Alteration (M=6.548, SD=2.535, $p$\textless.001). Color Alteration (M=5.305, SD=2.388) also has less Head-Turning than Size Alteration (M=6.548, SD=2.535, $p$=.002).
}
% ---------------------AR - Head-turning count - Position ------------------

% \begin{figure}[t]
% \centering
% \includegraphics[angle=270, width=.47\textwidth]{Figure/Final/AR_Turning_Count_Position.pdf}
% \caption{
% AR Experiment: Head-turning count result with the manipulated object's position.
% } 
% \label{figure:ar_turning_position}   
% % \vspace{-0.1cm}
% \end{figure}

% \begin{table}[ht] %
% \caption{AR Experiment: Statistic Head-turning counts result with the manipulated object's position.}
% \begin{tabular} {m{2.3cm} | m{1.1cm} | m{1cm}| m{0.8cm} | m{0.8cm}}
% \toprule 
%     {\textbf{Factor}} & {\textbf{$DOF$}}  & {\textbf{$F$}} & {\textbf{$p$}} & {\textbf{$\eta$\textsubscript{$p$}\textsuperscript{$2$}}}\\ \midrule
%     Level  & {3, 63} &  {3.426}  & {.022} & {.140} \\
%     Place  & {11, 231} &  {.890}  & {.551} & {.041} \\
%     \bottomrule
% \end{tabular}
% \label{table:ar_turning_position}
% %\vspace{-0.6cm}
% \end{table}

% % Different from the VR experiment result, 
% The Head-Turning count has a main effect on the floor of manipulated object located (F(3, 63)=3.426, $p$=.022, \textbf{$\eta$\textsubscript{$p$}\textsuperscript{$2$}}=.140). The 4th (M=3.09, SD=2.07) and 3rd (M=3.37, SD=1.28) floors' objects made less number of head-turning than the 1st floor's (M=4.22, SD=1.49, $p$=.039 for 4th; $p$=.014 for 3rd).

\subsubsection{Qualitative Analysis}

% Same as the VR experiment, we asked experience questionnaires after the end of each visual attention-disrupting condition.
There is a significant difference in difficulty to finding the altered objects between the visual attention-disrupting conditions(F=(1,42) = 5.988, $p$ = .004, \textbf{$\eta$\textsubscript{$p$}\textsuperscript{$2$}}=.160). 
The participants responded that Head-Turning 60$^\circ$ (M=4.32) and 30$^\circ$ (M=4.09) are more difficult than Flickering (M=2.95).
% They scored the difficulty for each visual attention-disrupting condition as 2.95 for flickering, 4.32 for 60\textdegree head-turning, and 4.09\textdegree head-turning condition(F=(1,42) = 5.988, $p$ = .004, \textbf{$\eta$\textsubscript{$p$}\textsuperscript{$2$}}=.160) from 1 (easiest) to 7 (hardest).
% The comments about the difficulty are similar to the VR experiment's result.
They mentioned that the reason why Head-Turning is more difficult is the movement makes them hard to focus on one part. P1 stated that \textit{``The flickering method was easier for me because it didn't require me to think as hard or keep as much stored in my short term memory at a time. I could just take in a few objects at a time, wait to see if they changed at all, then move on with confidence that I didn't miss the changing object.''}. P4 also mentioned \textit{``I had to refocus my attention after I turned back on the object I was looking at specifically whereas with flickering I could easily compare the two different shelves while maintaining my focus on the same location.''}
However, as inferred from the score of both Head-Turning conditions was not significantly different, the participants did not mention the difference in angle specifically.

Among the object alteration conditions, the participants stated that Removal was the easiest condition. For the Flicking method, 13 participants chose the Removal condition. for both the 60$^\circ$ Head-Turning, and 30$^\circ$ Head-Turning conditions, 14 participants chose the Removal condition. \modified{Conversely, the Size Alternation condition is responded to as the most challenging.}

%Participants' general answers for the search strategy were the same within the VR experiment. Regardless of the type of immersive environment, they had to focus from top to bottom and row by row within both visual attention-disrupting conditions.

%Here are some participants' comments on this AR experiment.%search strategy and conditions.

% \begin{quote}

% \textit{P5 - Strategy: ``I would look at each row and stare at it to memorize what the different objects shapes and colors where and then moved to the next one when there was no difference. ''}

% \textit{P8 - Strategy: ``looking at either 1 or 2 at a time and going down the row''}

% \textit{P4: ``I had to refocus my attention after I turned back on the object I was looking at specifically whereas with flickering I could easily compare the two different shelves while maintaining my focus on the same location.''}
% \end{quote}

% \input{Contents/Result_VR_AR.tex}

\section{Discussion}
% The results of our two user studies provide insights into the participants' performance and perception in different experimental conditions in VR and AR. 
The two user studies in VR and AR provide insights into the participants' performance and perception. In this section, we discuss our findings and compare the results between VR and AR \modified{using the between-subject design}.

% \subsection{
\paragraph{\textbf{Detection Time}}
% ----------- RQ - 1: FoV, VR / AR
The participants demonstrate faster detection of the altered object in the Flickering condition compared to the Head-Turning condition in both VR and AR. This result represents that when object alternation has occurred outside of FoV, it takes more time for participants to detect the changes compared to when it occurred within FoV. The participants' comments also support this, as they mentioned difficulties in paying attention to the objects when the objects moved out of the FoV.
This finding supports H1 that the participants would spend more time turning their heads and refocusing the scene leading the slower detection in the Head-Turning condition.
% Interestingly, in AR, there was no significant difference in detection time between the two different head-turning conditions, which rejects H3.

% this could be a limitation, we only 
In the AR experiment, different angles of the Head-Turning conditions do not affect the participants' detection time.
%This result shows that the change detection speed is not affected by how long the altered object was outside of the FoV. 
This also indicates that the narrower FoV does not significantly impact the detection of changes occurring outside of the FoV.
This rejects H4. However, it would be affected if the altered object remains outside of the FoV significantly longer period than just during the Head-Turning.
Further studies are needed to evaluate the effect of the significantly long period when the altered object is out of the FoV.
% that a larger angle with the head-turning condition has a worse effect on the change detection.

% ----------- RQ - 2: Alternation conditions
Regarding the object alteration conditions, our findings in both VR and AR align with the previous studies conducted in a 2D environment\cite{rensink1997see,ma2013change}.
% However, it contrasts with the recent study conducted in a VR environment, which reported no significant effect between the object alternation conditions \cite{martin2023study}.
% However, it contrasts with the recent study conducted in a VR environment \cite{martin2023study}.
With the Removal condition, the detection time is significantly faster than the result with the \modified{Color Alteration and Size Alteration conditions in both VR and AR.}
This finding supports H2 that the participants would find the altered object with the Removal condition easily. 

% In the AR environment, the result shows that the detection time with the removal condition is also significantly faster than the result with other conditions.
In AR, however, the Size Alteration condition has significantly slower detection time than the Removal and Color Alteration conditions.
Interestingly, the increase in detection time with the Color Alternation condition in AR is only 5.7\% compared to VR, while the other conditions show increases of at least 26.7\% and up to 29.0\%.
This result suggests that the perception of the color visual channel is not significantly decreased in AR compared to VR. This supports H3, that the visual channel would be perceived differently in AR and VR.
Additionally, this result aligns with the conventional approach of visual perception (i.e., the coherence theory and triadic architecture) \cite{rensink2000visual,rensink2002change} which proposes each visual channel is perceived differently.
Consequently, these detection time results represent that the limited FoV and type of visual channel influence how fast the observer can detect changes.

The 3-way mixed ANOVA (2 immersive environments \modified{(between-subject)} $\times$ 2 visual attention-disturbing\footnote{For the mixed ANOVA, we only include the 60$^\circ$ Head-Turning from the AR study.} $\times$ 3 object alteration) shows a main effect on the detection time of the immersive environment (F(1, 42)=5.051, $p$=.030, $\eta$\textsubscript{$p$}\textsuperscript{$2$}=.107). However, no interaction effect is found related to the immersive environment. Overall, the average detection time in the VR environment (M=14.97s, SD=6.013) is significantly faster than the one in the AR environment (M=17.89s, SD=7.981).
This difference can be because of the different sizes of the FoV \modified{that can occur the effect from the vertical displacement,} and different environmental characteristics, such as different object transparency in VR and AR environments.

\paragraph{\textbf{Detection Rate}}
These detection rate results indicate which conditions are more likely to cause confusion for observers during change detection.
% ----------- RQ - 1: FoV, VR / AR
% p = .052
% In the VR environment, an interaction effect is observed between the visual attention-disturbing and the object alteration conditions.
% In the Head-Turning condition, the Removal condition's detection rate is higher than the results of \modified{the Size Alteration condition.}
% This indicates that the size alternation conditions can lead to confusion during change detection when combined with the Head-Turning condition, especially when the observer cannot maintain focus on the same location.
% This finding supports H1 and H2, indicating that when moving the object out of the FoV can cause confusion for each visual channel differently.
% ----------- RQ - 2: Alternation conditions
 In both immersive environments, the Removal condition has higher detection rates than the Size Alteration condition.
In VR, the Color Alteration condition has significantly lower detection rates than the Removal condition. However, in AR, it's significantly higher than the results from the Size Alteration conditions. 
This supports H1 and suggests that the color perception in both immersive environments is not significantly different.

The mixed ANOVA does not reveal a significant main effect on the detection rate between VR and AR. %The average detection rate result in the VR and AR experiments does not show a significant difference.
This suggests that, despite potential differences in visual perception speed between VR and AR, the accuracy of perception remains similar in both VR and AR.
\paragraph{\textbf{Timeout Rate}}
Different from the detection rate, this timeout rate results represent the ratio of tasks in which the participants were not able to find the altered object despite their awareness that one of the objects has been altered. This highlights instances where participants experienced difficulty in identifying the altered object within the given time limit (60 seconds).
% ----------- RQ - 1: FoV, VR / AR
Similar to the detection time and rate, the participants find the altered object faster when it is changed inside of FoV in both types of immersive environments.
In VR and AR, the Flickering condition has a significantly lower timeout rate than the Head-Turning condition.
% This result represents that the change blindness effect has easily occurred when the object is changed outside of FoV, and it is a consistent result with the previous change blindness study in VR \cite{martin2023study}, and it supports H1.
\modified{
In the VR environment, an interaction effect is observed between the visual attention-disturbing and the object alteration conditions.
 Within the Head-Turning condition, the Removal condition has a lower timeout rate than the Size Alteration condition. Furthermore, under the Size Alteration condition, the timeout rate for the Flickering condition is lower than for the Head-Turning condition.
}
This result indicates that change blindness is more likely to occur when an object changes outside the FoV. In VR, the Size Alteration is particularly more influenced by object changes outside the FoV compared to the Removal condition. Consequently, it supports H1.
Additionally, in AR, the different angles of the Head-Turning conditions do not affect the result.
It is the same with the detection time results and rejects H4; the Head-Turning angle does not affect the timeout rate.

% ----------- RQ - 2: Alternation conditions
\modified{The Removal condition results in a significantly lower timeout rate than the Size Alteration condition in both environments. However, only in AR does the Removal condition yield a lower result than that of the Color Alteration condition.}
% The Removal condition makes significantly less Timeout Rate than other object alternation conditions only in AR.
However, although the \modified{effect of Color Alternation conditions in VR is not significant, the order of the timeout rate for the condition is the same within AR.}
From this, we can guess that the AR environment has a possibility to reinforce the \modified{Color} Alternation condition's effect.

The previous study on change blindness in VR conducted by Martin et al. \cite{martin2023study} reported the effect of FoV on the timeout rate. This is in line with our findings, where changes outside the FoV are more difficult to detect than those within the FoV. However, Martin et al. did not find any significant effects based on the type of object alternation. Their experiments were conducted in an environment that closely resembled daily visual experiences, populated with a variety of object shapes that might divert the participant's attention. This might account for the differences observed in our results.

\paragraph{\textbf{Head-Turning Count}}
This Head-Turning count represents the number of times the participants need to rotate their heads to detect a change that occurs outside of the FoV.

In both VR and AR, the Removal condition required the fewest Head-Turning counts to detect change.
This finding indicates the Removal condition is more easily detected with fewer alterations, aligning with the results of detection time and rate. This result supports H1 that the presence or absence of objects is the most memorable aspect of the visual stimulus.

In VR, the required counts for different object alternation conditions are similar to one another, but this is not the case in AR.
This difference may come from the mixed entities between virtual and real entities in AR.

\subsection{Comparison to the previous studies}
Previous studies in immersive environments designed experiments using objects of various shapes, backgrounds, and distances ~\cite{steinicke2011change, martin2023study}.
Such variations can lead to results that are influenced by a mix of the environment and object characteristics, such as shape, color, size, and position.
% However, Martin et al. ~\cite{martin2023study} had a different result from ours.
Even though Martin et al. \cite{martin2023study} did their study in a VR setting, they were not able to identify any effects related to object alternation types.  They claimed that their results differed from the previous change blindness study in some aspects that concentrated on 2D image alternation types within a traditional desktop environment ~\cite{ma2013change}. Understanding this gap and the difficulties in drawing clear conclusions from complex settings, our study aimed to set up a more controlled environment to better analyze the factors involved.  Consequently, our findings highlight distinct effects across various alteration types and offer a comparative analysis between VR and AR. The controlled setting helped in minimizing confounding variables, thereby providing a clearer insight into the factors at play in immersive environments. 

% ------------------------ 
% Explain what points are better than previous studies
%

\subsection{Limitation}
In this section, we discuss the limitations of our work. One limitation may arise from the limitation of the devices and their FoVs. Our study is conducted in VR and AR environments and compared results within both environments.
However, due to the limitation of the devices and their FoVs, we were not able to fully control all variables, such as FoV, interaction techniques (controller in VR and hand gesture in AR), and transparency of virtual objects. Further studies are needed to evaluate the effect of the significantly long period when the altered object is out of the FoV.

\modified{
The second limitation stems from the challenges in accurately measuring the duration of a user's head-turning. We did not explicitly record the actual duration of head-turning as it was challenging to separate the duration of head-turning and the time to focus on the scene as these would be expected to be mixed. During the experiment, participants were free to move their heads vertically and even horizontally while they found the changed object. In the Flickering condition, the object was altered periodically for a constant time. However, in the Head-Turning condition, such alteration period was handled by participants, leading to potential variability. A more precise measurement of head-turning duration could provide more accurate results regarding the effect of different visual attention-disrupting conditions on detection time.  %Therefore, distinguishing between the time spent turning their head and the time spent merely focusing is challenging. Our reported completion time does not %It was difficult to distinguish this participant’s motion between turning their head for the object alteration or just moving their head.
% Within the Flickering condition, the object was altered periodically for a constant time, but within the Head-Turning condition, that alteration period was handled by participants. Additionally, participants could move their heads vertically and even horizontally during the task.
% It was problematic to distinguish this participant's motion between turning their head for the object alternation or just moving their head.
% This makes a difference in altering time, making an inappropriateness to compare the two conditions' results directly.
% During the task, participants could move their heads vertically and even horizontally. Within the Head-Turning condition, this free action made measuring the exact time between their head-turning hard.
}
% The detection time is a common measurement for the change blindness study.
% However, our result shows that the participant's search strategy dominantly affects the detection time and, particularly, the spatial searching pattern.
% Even though we had spatial balancing to ignore that limitation, that issue was only removed statistically.

\modified{
The third limitation pertains to the age groups of our study's participants. According to related research, visual memory and perception are reduced by age \cite{naveh2000adult, ko2014understanding}. The age range of participants in our study was focused from 18 to 24. Therefore, outcomes with other age-ranged participant groups might differ from our findings.
}

Finally, most change blindness studies, even ours, have repeated measuring experiments that ask to find some changed thing. However, in that procedure, participants will focus on objects and try to remember as much as they can.
Consequently, this kind of study can be just a visual memory test even if the visual working memory is related to the change blindness phenomenon.
% \textbf{Habituation effects}

%\textbf{future work}
\section{Conclusion}

In this paper, we examine the effects of the limited FoV and the object alternation conditions for change blindness within controlled VR and AR environments. Our results indicate that a limited FoV impacts both the detection time and rates of change detection. Specifically, when object alternations occur outside the FoV, detecting change blindness becomes significantly more challenging than when they occur within the FoV. Additionally, our findings reveal that different object alternation conditions influence the user's ability to detect changes. Specifically, in both VR and AR environments, it is significantly easier to detect changes when an object alternates by appearing and disappearing than when its size or color is altered. Our results also highlight differences in change blindness detection between VR and AR environments. Detecting changes is generally easier in VR than in AR, a difference that might be attributed to factors like variations in FoV and the contrast between real and virtual backgrounds.

% In the VR environment, change detection is easier than in AR, but it can be because of a different size of FoV, transparency of virtual objects, or different types of background (real and virtual).

%From these results, we can understand the impact of FoV, the type of immersive environment, and the object alternation condition on the change blindness phenomenon.

%\bibliographystyle{abbrv}
\bibliographystyle{abbrv-doi}

\bibliography{reference}

\begin{thebibliography}{10}

\bibitem{ahissar1996learning}
M.~Ahissar and S.~Hochstein.
\newblock Learning pop-out detection: Specificities to stimulus
  characteristics.
\newblock {\em Vision research}, 36(21):3487--3500, 1996.

\bibitem{baudisch2003halo}
P.~Baudisch and R.~Rosenholtz.
\newblock Halo: a technique for visualizing off-screen objects.
\newblock In {\em Proceedings of the SIGCHI conference on Human factors in
  computing systems}, pp. 481--488, 2003.

\bibitem{bays2009precision}
P.~M. Bays, R.~F. Catalao, and M.~Husain.
\newblock The precision of visual working memory is set by allocation of a
  shared resource.
\newblock {\em Journal of vision}, 9(10):7--7, 2009.

\bibitem{beanland2017change}
V.~Beanland, A.~J. Filtness, and R.~Jeans.
\newblock Change detection in urban and rural driving scenes: Effects of target
  type and safety relevance on change blindness.
\newblock {\em Accident Analysis \& Prevention}, 100:111--122, 2017.

\bibitem{brady2011review}
T.~F. Brady, T.~Konkle, and G.~A. Alvarez.
\newblock A review of visual memory capacity: Beyond individual items and
  toward structured representations.
\newblock {\em Journal of vision}, 11(5):4--4, 2011.

\bibitem{castillo2020allocentric}
J.~Castillo~Escamilla, J.~J. Fern{\'a}ndez~Castro, S.~Baliyan, J.~J.
  Ortells-Pareja, J.~J. Ortells~Rodr{\'\i}guez, and J.~M. Cimadevilla.
\newblock Allocentric spatial memory performance in a virtual reality-based
  task is conditioned by visuospatial working memory capacity.
\newblock {\em Brain Sciences}, 10(8):552, 2020.

\bibitem{charlton2013driving}
S.~G. Charlton and N.~J. Starkey.
\newblock Driving on familiar roads: Automaticity and inattention blindness.
\newblock {\em Transportation research part F: traffic psychology and
  behaviour}, 19:121--133, 2013.

\bibitem{cummings2016immersive}
J.~J. Cummings and J.~N. Bailenson.
\newblock How immersive is enough? a meta-analysis of the effect of immersive
  technology on user presence.
\newblock {\em Media psychology}, 19(2):272--309, 2016.

\bibitem{davies2007change}
G.~Davies and S.~Hine.
\newblock Change blindness and eyewitness testimony.
\newblock {\em The Journal of psychology}, 141(4):423--434, 2007.

\bibitem{divita2004verification}
J.~DiVita, R.~Obermayer, W.~Nugent, and J.~M. Linville.
\newblock Verification of the change blindness phenomenon while managing
  critical events on a combat information display.
\newblock {\em Human factors}, 46(2):205--218, 2004.

\bibitem{fitzgerald2016change}
R.~J. Fitzgerald, C.~Oriet, and H.~L. Price.
\newblock Change blindness and eyewitness identification: Effects on accuracy
  and confidence.
\newblock {\em Legal and Criminological Psychology}, 21(1):189--201, 2016.

\bibitem{gaffary2017ar}
Y.~Gaffary, B.~Le~Gouis, M.~Marchal, F.~Argelaguet, B.~Arnaldi, and
  A.~L{\'e}cuyer.
\newblock Ar feels “softer” than vr: Haptic perception of stiffness in
  augmented versus virtual reality.
\newblock {\em IEEE transactions on visualization and computer graphics},
  23(11):2372--2377, 2017.

\bibitem{gaspar2013change}
J.~G. Gaspar, M.~B. Neider, D.~J. Simons, J.~S. McCarley, and A.~F. Kramer.
\newblock Change detection: training and transfer.
\newblock {\em PloS one}, 8(6):e67781, 2013.

\bibitem{jones2008effects}
J.~A. Jones, J.~E. Swan, G.~Singh, E.~Kolstad, and S.~R. Ellis.
\newblock The effects of virtual reality, augmented reality, and motion
  parallax on egocentric depth perception.
\newblock In {\em Proceedings of the 5th symposium on Applied perception in
  graphics and visualization}, pp. 9--14, 2008.

\bibitem{ko2014understanding}
P.~C. Ko, B.~Duda, E.~Hussey, E.~Mason, R.~J. Molitor, G.~F. Woodman, and B.~A.
  Ally.
\newblock Understanding age-related reductions in visual working memory
  capacity: Examining the stages of change detection.
\newblock {\em Attention, Perception, \& Psychophysics}, 76:2015--2030, 2014.

\bibitem{levin1997failure}
D.~T. Levin and D.~J. Simons.
\newblock Failure to detect changes to attended objects in motion pictures.
\newblock {\em Psychonomic Bulletin \& Review}, 4(4):501--506, 1997.

\bibitem{ma2013change}
L.-Q. Ma, K.~Xu, T.-T. Wong, B.-Y. Jiang, and S.-M. Hu.
\newblock Change blindness images.
\newblock {\em IEEE transactions on visualization and computer graphics},
  19(11):1808--1819, 2013.

\bibitem{martin2023study}
D.~Martin, X.~Sun, D.~Gutierrez, and B.~Masia.
\newblock A study of change blindness in immersive environments.
\newblock {\em IEEE Transactions on Visualization and Computer Graphics}, 2023.

\bibitem{naveh2000adult}
M.~Naveh-Benjamin.
\newblock Adult age differences in memory performance: tests of an associative
  deficit hypothesis.
\newblock {\em Journal of Experimental Psychology: Learning, Memory, and
  Cognition}, 26(5):1170, 2000.

\bibitem{nelson2011change}
K.~J. Nelson, C.~Laney, N.~B. Fowler, E.~D. Knowles, D.~Davis, and E.~F.
  Loftus.
\newblock Change blindness can cause mistaken eyewitness identification.
\newblock {\em Legal and criminological psychology}, 16(1):62--74, 2011.

\bibitem{pertzov2012forgetting}
Y.~Pertzov, M.~Y. Dong, M.-C. Peich, and M.~Husain.
\newblock Forgetting what was where: The fragility of object-location binding.
\newblock {\em PLoS One}, 7(10):e48214, 2012.

\bibitem{piumsomboon2017covar}
T.~Piumsomboon, Y.~Lee, G.~Lee, and M.~Billinghurst.
\newblock Covar: a collaborative virtual and augmented reality system for
  remote collaboration.
\newblock In {\em SIGGRAPH Asia 2017 Emerging Technologies}, pp. 1--2. 2017.

\bibitem{potter2014detecting}
M.~C. Potter, B.~Wyble, C.~E. Hagmann, and E.~S. McCourt.
\newblock Detecting meaning in rsvp at 13 ms per picture.
\newblock {\em Attention, Perception, \& Psychophysics}, 76:270--279, 2014.

\bibitem{rensink2000dynamic}
R.~A. Rensink.
\newblock The dynamic representation of scenes.
\newblock {\em Visual cognition}, 7(1-3):17--42, 2000.

\bibitem{rensink2000visual}
R.~A. Rensink.
\newblock Visual search for change: A probe into the nature of attentional
  processing.
\newblock {\em Visual cognition}, 7(1-3):345--376, 2000.

\bibitem{rensink2002change}
R.~A. Rensink.
\newblock Change detection.
\newblock {\em Annual review of psychology}, 53(1), 2002.

\bibitem{rensink2002failure}
R.~A. Rensink.
\newblock Failure to see more than one change at a time.
\newblock {\em Journal of vision}, 2(7):245--245, 2002.

\bibitem{rensink1997see}
R.~A. Rensink, J.~K. O'regan, and J.~J. Clark.
\newblock To see or not to see: The need for attention to perceive changes in
  scenes.
\newblock {\em Psychological science}, 8(5):368--373, 1997.

\bibitem{romer2014adolescence}
D.~Romer, Y.-C. Lee, C.~C. McDonald, and F.~K. Winston.
\newblock Adolescence, attention allocation, and driving safety.
\newblock {\em Journal of Adolescent Health}, 54(5):S6--S15, 2014.

\bibitem{sauer2022assessment}
Y.~Sauer, A.~Sipatchin, S.~Wahl, and M.~Garc{\'\i}a~Garc{\'\i}a.
\newblock Assessment of consumer vr-headsets’ objective and subjective field
  of view (fov) and its feasibility for visual field testing.
\newblock {\em Virtual Reality}, 26(3):1089--1101, 2022.

\bibitem{simons1999gorillas}
D.~J. Simons and C.~F. Chabris.
\newblock Gorillas in our midst: Sustained inattentional blindness for dynamic
  events.
\newblock {\em perception}, 28(9):1059--1074, 1999.

\bibitem{simons1997change}
D.~J. Simons and D.~T. Levin.
\newblock Change blindness.
\newblock {\em Trends in cognitive sciences}, 1(7):261--267, 1997.

\bibitem{simons2005change}
D.~J. Simons and R.~A. Rensink.
\newblock Change blindness: Past, present, and future.
\newblock {\em Trends in cognitive sciences}, 9(1):16--20, 2005.

\bibitem{smart2019measuring}
S.~Smart and D.~A. Szafir.
\newblock Measuring the separability of shape, size, and color in scatterplots.
\newblock In {\em Proceedings of the 2019 CHI Conference on Human Factors in
  Computing Systems}, pp. 1--14, 2019.

\bibitem{steinicke2011change}
F.~Steinicke, G.~Bruder, K.~Hinrichs, and P.~Willemsen.
\newblock Change blindness phenomena for virtual reality display systems.
\newblock {\em IEEE Transactions on visualization and computer graphics},
  17(9):1223--1233, 2011.

\bibitem{suchow2011motion}
J.~W. Suchow and G.~A. Alvarez.
\newblock Motion silences awareness of visual change.
\newblock {\em Current Biology}, 21(2):140--143, 2011.

\bibitem{suma2010exploiting}
E.~A. Suma, S.~Clark, S.~L. Finkelstein, and Z.~Wartell.
\newblock Exploiting change blindness to expand walkable space in a virtual
  environment.
\newblock In {\em 2010 IEEE Virtual Reality Conference (VR)}, pp. 305--306.
  IEEE, 2010.

\bibitem{suma2011leveraging}
E.~A. Suma, S.~Clark, D.~Krum, S.~Finkelstein, M.~Bolas, and Z.~Warte.
\newblock Leveraging change blindness for redirection in virtual environments.
\newblock In {\em 2011 IEEE Virtual Reality Conference}, pp. 159--166. IEEE,
  2011.

\bibitem{triesch2003you}
J.~Triesch, D.~H. Ballard, M.~M. Hayhoe, and B.~T. Sullivan.
\newblock What you see is what you need.
\newblock {\em Journal of vision}, 3(1):9--9, 2003.

\bibitem{ungerleider1998neural}
L.~G. Ungerleider, S.~M. Courtney, and J.~V. Haxby.
\newblock A neural system for human visual working memory.
\newblock {\em Proceedings of the National Academy of Sciences},
  95(3):883--890, 1998.

\bibitem{ware2019information}
C.~Ware.
\newblock {\em Information visualization: perception for design}.
\newblock Morgan Kaufmann, 2019.

\bibitem{white2010blind}
C.~B. White and J.~K. Caird.
\newblock The blind date: The effects of change blindness, passenger
  conversation and gender on looked-but-failed-to-see (lbfts) errors.
\newblock {\em Accident Analysis \& Prevention}, 42(6):1822--1830, 2010.

\end{thebibliography}
\end{document}